\documentclass{aa}

\usepackage[varg]{txfonts}

\usepackage{graphicx}	
\usepackage{amsmath}	
\usepackage{subfigure}
\usepackage{epstopdf}
\usepackage[colorlinks, linkcolor=blue, urlcolor=blue, citecolor=blue]{hyperref}

\usepackage{booktabs}
\usepackage{ulem}

\usepackage{threeparttable}
\usepackage{enumitem}
\usepackage{longtable}

\def\be{\begin{equation}}
\def\ee{\end{equation}}

\usepackage{CJK}
\usepackage{upgreek}

\begin{document}

\title{Narrow spectra of repeating fast radio bursts: A magnetospheric origin
}

\begin{CJK*}{UTF8}{gbsn}
\author{Wei-Yang Wang (王维扬)\inst{\ref{inst1},\ref{inst2},\ref{inst3}}
\and Yuan-Pei Yang (杨元培)\inst{\ref{inst4},\ref{inst5}}
\and Hong-Bo Li (李洪波)\inst{\ref{inst3},\ref{inst6}}
\and Jifeng Liu (刘继峰)\inst{\ref{inst1},\ref{inst7},\ref{inst8}}
\and Renxin Xu (徐仁新)\inst{\ref{inst2},\ref{inst3},\ref{inst6}}
}

\institute{School of Astronomy and Space Science, University of Chinese Academy of Sciences, Beijing 100049, PR China \label{inst1}
\\
\email{wywang@ucas.ac.cn}
\and
State Key Laboratory of Nuclear Physics and
Technology, School of Physics, Peking University, Beijing 100871, PR China \label{inst2}
\and
Department of Astronomy, School of Physics, Peking University, Beijing 100871, PR China \label{inst3}
\and
South-Western Institute for Astronomy Research, Yunnan University, Kunming, Yunnan 650504, PR China \label{inst4}
\and 
Purple Mountain Observatory, Chinese Academy of Sciences, Nanjing 210023, PR China \label{inst5}
\and
Kavli Institute for Astronomy and Astrophysics, Peking University, Beijing 100871, PR China \label{inst6}
\and
New Cornerstone Science Laboratory, National Astronomical Observatories, Chinese Academy of Sciences, Beijing 100101, PR China \label{inst7}
\and
Institute for Frontiers in Astronomy and Astrophysics, Beijing Normal University, Beijing 102206, PR China \label{inst8}
}

\date{Received XXX / Accepted XXX}

\abstract{
Fast radio bursts (FRBs) can present a variety of polarization properties, and some of them have narrow spectra.
We study spectral properties from perspectives of intrinsic radiation mechanisms and absorption during the waves propagating in the magnetosphere.
The intrinsic radiation mechanisms are considered by invoking quasi-periodic bunch distribution and perturbations on charged bunches moving on curved trajectories.
The narrow-band emission likely reflects some quasi-periodic structure on the bulk of bunches, which may be due to quasi-periodically sparking in a ``gap'' or quasi-monochromatic Langmuir waves.
A sharp spike would appear at the spectrum if the perturbations can induce a monochromatic oscillation of bunches, however, it is hard to create a narrow spectrum because the Lorentz factor has large fluctuations so that the spike disappears.
Both the bunching mechanism and perturbations scenarios share the same polarization properties with a uniformly distributed bulk of bunches.
We investigate absorption effects including Landau damping and curvature self-absorption in the magnetosphere, which are significant at low frequencies.
Subluminous O-mode photons can not escape from the magnetosphere due to the Landau damping, leading to a height-dependent lower frequency cut-off.
Spectra can be narrow when the frequency cut-off is close to the characteristic frequency of curvature radiation, while such conditions can only be met sometimes.
The spectral index is 5/3 at low-frequency bands due to the curvature self-absorption but not as steep as the observations.
The intrinsic radiation mechanisms are more likely to generate the observed narrow spectra of FRBs rather than the absorption effects.
}


\titlerunning{}
\authorrunning{Wang, W.-Y., et al.}

\maketitle

\section{Introduction}\label{sec1}
\end{CJK*}
Fast radio bursts (FRBs) are millisecond-duration and energetic radio pulses (see \citealt{2019ARA&A..57..417C,2019A&ARv..27....4P}, for reviews).
They are similar in some respects to single pulses from radio pulsars, however, the bright temperature of FRBs is many orders of magnitude higher than that of normal pulses from Galactic pulsars.
The physical origin(s) of FRB are still unknown but the only certainty is that their radiation mechanisms must be coherent.

Magnetars, as the most likely candidate, have been invoked to interpret the emission of FRBs, especially repeating ones.
The models can be generally divided into two categories by considering the distance of the emission region from the magnetar \citep{2020Natur.587...45Z}: emission within the magnetosphere (pulsar-like model, e.g., \citealt{2014PhRvD..89j3009K,2017MNRAS.468.2726K,2018ApJ...868...31Y,2019ApJ...879....4W,2020MNRAS.498.1397L,2021MNRAS.508L..32C,2022ApJ...927..105W,2022ApJ...925...53Z,2023ApJ...958...35L,2023MNRAS.522.2448Q}), and emission from relativistic shocks region far outside the magnetosphere (GRB-like model, e.g., \citealt{2019MNRAS.485.4091M,2020ApJ...896..142B,2020ApJ...899L..27M,2022arXiv221013506C,2022ApJ...927....2K}).
The magnetar origin for at least some FRBs was established after the detection of FRB 20200428D, \footnote{FRB 20200428D was regarded as FRB 20200428A previously until other three burst events detected \citep{2023arXiv231016932G}.}
an FRB-like burst from SGR J1935+2154, which is a Galactic magnetar \citep{2020Natur.587...59B,2020Natur.587...54C}.
Nevertheless, the localization of FRB 20200120E to an old globular cluster in M81 \citep{2021ApJ...910L..18B,2021ApJ...919L...6M,2022Natur.602..585K} challenges that a young magnetar born via a massive stellar collapse, which means that the magnetar formed in accretion-induced collapse or a system associated with a compact binary merger \citep{2021ApJ...917L..11K,2022MNRAS.510.1867L}.

Spectral time characteristics of FRBs are significant tools to gain a deeper understanding of emission mechanisms.
The burst morphologies of FRBs are varied.
Some FRBs have narrow bandwidth, for instance, the sub-bursts of FRB 20220912A show that the relative spectral bandwidth of the radio bursts was distributed near at $\Delta\nu/\nu\simeq0.2$ \citep{2023ApJ...955..142Z}.
Similar phenomena of narrow-band can appear at FRB 20121102A, in which spectra are modeled by simple power laws with indices ranging from $-10$ to $+14$ \citep{2016Natur.531..202S}.
Simple narrow-band and other three features (simple broadband, temporally complex, and downward drifting) are summarized as four observed archetypes of burst morphology among the first CHIME/FRB catalog \citep{2021ApJ...923....1P}.
The narrow spectra may result from small deflection angle radiation, coherent processes or modulated by scintillation and plasma lensing in the FRB source environment \citep{2023ApJ...956...67Y}.

Downward drift in the central frequency of consecutive sub-bursts with later-arriving time, has been discovered in at least some FRBs \citep{2019ApJ...885L..24C,2019ApJ...876L..23H,2020ApJ...891L...6F}.
Besides, a variety of drifting patterns including upward drifting, complex structure, and no drifting subpulse, was found in a sample of more than 600 bursts detected from the repeater source FRB 20201124A \citep{2022RAA....22l4001Z}.
Such a downward drifting structure was unprecedentedly seen in pulsar PSR B0950+08 \citep{2022A&A...658A.143B}, suggesting that FRBs originate from the magnetosphere of a pulsar.
Downward drifting pattern can be well understood by invoking magnetospheric curvature radiation \citep{2019ApJ...876L..15W}, however, it is unclear why the spectral bandwidth can be extremely narrow, i.e., $\Delta\nu/\nu\simeq0.1$.

Polarization properties carry significant information about radiation mechanisms to shed light on the possible origin of FRBs.
High levels of linear polarization are dominant for most sources and most bursts for individual repeating sources (e.g., \citealt{2018Natur.553..182M,2020MNRAS.497.3335D,2020Natur.586..693L,2023ApJ...957L...8S}), while some bursts have significant circular polarization fractions, which can be even up to $90\%$ (e.g., \citealt{2015Natur.528..523M,2022Natur.611E..12X,2022RAA....22l4003J,2022MNRAS.512.3400K}).
The properties are reminiscent of pulsars, which show a wide variety of polarization fractions between sources.
The emission is highly linearly polarized when the line of sight (LOS) is confined into a beam angle, while highly circular polarization presents when LOS is outside, by considering bunched charges moving at ultra-relativistic speed in magnetosphere \citep{2022MNRAS.517.5080W,2023ApJ...943...47L}.

In this work, we attempt to understand the origin(s) of the observed narrow spectra of FRBs from the views of intrinsic radiation mechanisms and absorption effects in the magnetosphere.
The paper is organized as follows. 
The intrinsic radiation mechanisms by charged bunches in the magnetosphere are discussed in Section \ref{sec2}.
We investigate two scenarios of the spectrum, including the bunch mechanism (Section \ref{sec2.1}) and perturbations on the bunch moving at curved trajectories (Section \ref{sec2.2}).
The polarization properties of the intrinsic radiation mechanisms are summarized in Section \ref{sec2.3}.
In Section \ref{sec3}, we discuss several possible mechanisms as triggers that may generate emitting charged bunches (Section \ref{sec3.1}), and oscillation and Alfv\'en wave.
Some absorption effects during the wave propagating in the magnetosphere are investigated in Section \ref{sec5}.
We discuss the bunch's evolution in the magnetosphere in Section \ref{sec4}.
The results are discussed and summarized in Section \ref{sec6} and \ref{sec7}.
The convention $Q_x=Q/10^x$ in cgs units and spherical coordinates ($r,\,\theta,\,\varphi$) concerning the magnetic axis are used throughout the paper.

\section{Intrinsic Radiation Mechanism}\label{sec2}

\begin{figure*}
\centering
\includegraphics[width=17cm]{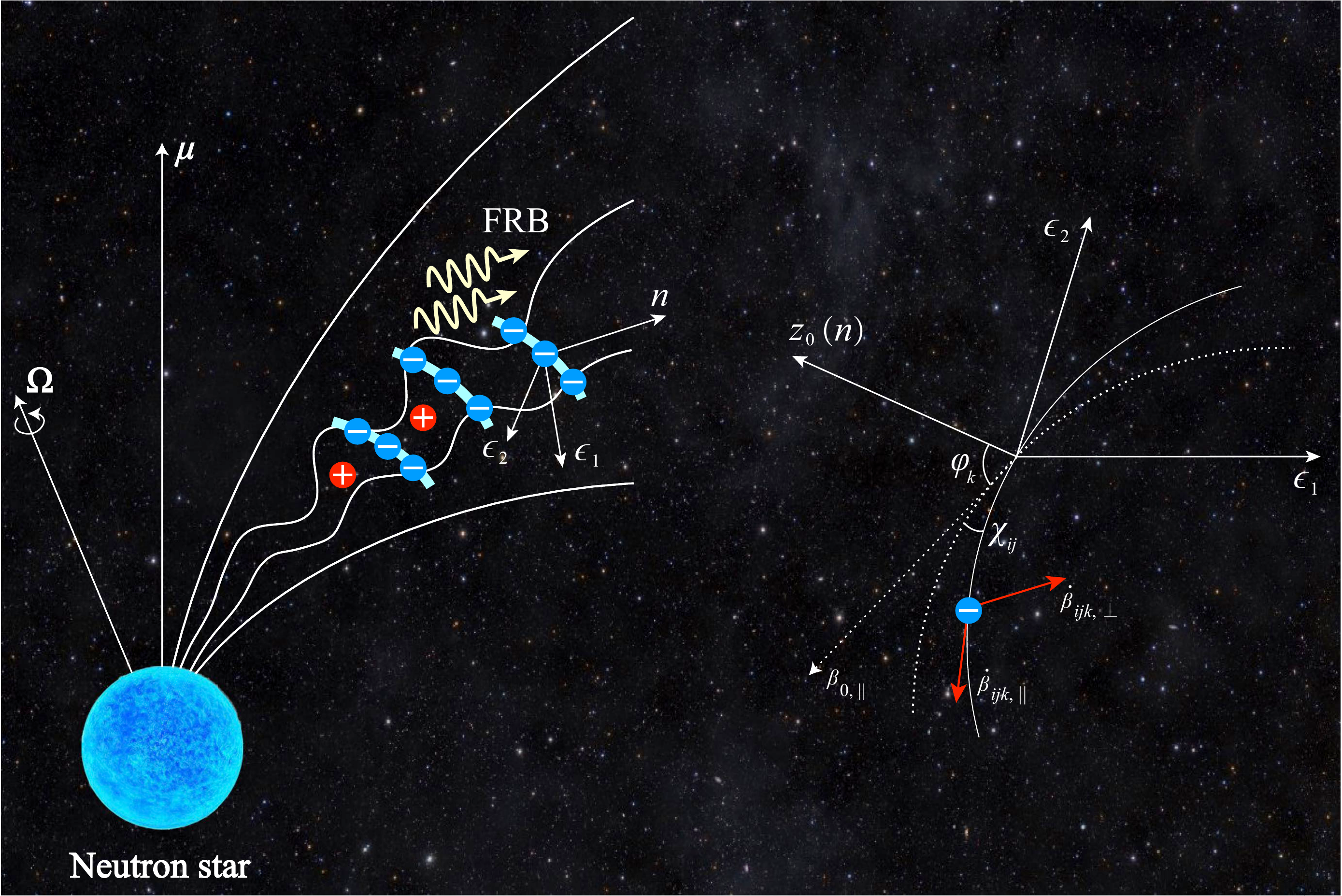}
\caption{Left: Schematic diagram of a bulk of bunches moving in the magnetosphere. The light
solid blue line shows the slice in which electrons emit roughly the same phase. The unit vector
of the LOS is denoted by $\boldsymbol{n}$, and $\boldsymbol{\epsilon}_1$ and $\boldsymbol{\epsilon}_2$ denote the two polarization components.
Right: Geometry for instantaneous circular motion for a particle identified by $ijk$ in a bunch.
The trajectory lies in the $\boldsymbol{\beta}_{0,\|}$-$\boldsymbol{\epsilon}_1$ plane. At the retarded time $t = 0$, the electron is at the origin. The dotted trajectory shows an electron at the origin.
The angle between the solid and dotted trajectory is $\chi_{ij}$ at $t = 0$. The angle between LOS ($\boldsymbol{z}_0$) and the trajectory plane is $\varphi_k$.
}
\label{fig:bunch}
\end{figure*}

A sudden trigger may happen on the stellar surface and sustain at least a few milliseconds to create free charges.
Such a trigger event is a sudden and violent process in contrast to a consecutive ``sparking'' process in the polar cap region of a pulsar \citep{1975ApJ...196...51R}.
The charges can form bunches, and the total emission of charges is coherently enhanced significantly if the bunch size is smaller than the half wavelength.
Note that electrons as the emitting charges are the point of interest in the following discussion, leading to negatively net-charged bunches.
We attempt to study how to trigger the emitting charges and how to form bunches in Section \ref{sec3}.

Within the magnetosphere, the charged particles are suddenly accelerated to ultra-relativistic velocities and stream outward along the magnetic field lines.
The charges hardly move across the magnetic field lines because the cyclotron cooling is very fast in a strong magnetic field and charges stay in the lowest Landau level so that charges' trajectories track with the field lines essentially.

Curvature radiation can be produced via the perpendicular acceleration for a charge moving at a curved trajectory.
It is a natural consequence that charges moving at curved trajectories, that has been widely discussed in pulsars \citep{1975ApJ...196...51R,1977ApJ...212..800C,2000ApJ...544.1081M,2004ApJ...600..872G,2021ApJ...911..152G} and FRBs \citep{2017MNRAS.468.2726K,2018MNRAS.481.2946K,2018MNRAS.477.2470L,2018A&A...613A..61G,2018ApJ...868...31Y,2022ApJ...927..105W,2023ApJ...956...35C}.
However, the noticeable difference is that there should be an electric field parallel ($E_\|$) to the local magnetic field $\boldsymbol{B}$ in the FRB emission region because the enormous emission power of FRB lets the bunches cooling extremely fast.
The $E_\|$ sustains the energy of bunches so that they can radiate for a long enough time to power FRBs.
The equation for motion in the FRB emission region in the lab frame can be written as
\be
N_{\mathrm{e}}E_\| e{\rm d}s-\mathcal{L}_{\mathrm{b}}{\rm d}t=N_{\mathrm{e}}m_{\rm e} c^2{\rm d}\gamma,
\label{eq:motion}
\ee
where $N_{\mathrm{e}}$ is the number of net charges in one bunch that are radiating coherently, $c$ is the speed of light, $e$ is the elementary charge, $m_{\mathrm{e}}$ is the positron mass, $\gamma$ is the Lorentz factor of the bunch, and $\mathcal{L}_{\mathrm{b}}$ is the luminosity of a bunch.
The timescale for the $E_\|$ that can be balanced by radiation damping is essentially the cooling timescale of the bunches, which is much shorter than the FRB duration.
As a result, bunches stay balanced throughout the FRB emission process.

Bunches in the magnetosphere are accelerated to ultra-relativistic velocity.
A stable $E_\|$ ($\partial E/\partial t=0$) may sustain the Lorentz factor of bunches to be a constant.
The emission of bunches is confined mainly in a narrow conal region due to the relativistic beaming effect, so the line of sight (LOS) sweeps across the cone within a very short time.
According to the properties of the Fourier transform, the corresponding spectrum of the beaming emission within a short time is supposed to be wide-band.
This is a common problem among the emission from relativistic charges.
In the following discussion, we attempt to understand the narrow-band FRB emission by considering the bunching mechanism (Section \ref{sec2.1}) and perturbation on the curved trajectories (Section \ref{sec2.2}).

\subsection{Bunching Mechanism}\label{sec2.1}
The waves are coherently enhanced significantly when the emitting electrons form charged bunches, which is the fundamental unit of coherent emission.
Different photon arrival delays are attributed to charges moving in different trajectories, so that
charges in the horizontal plane (a light solid blue line denotes a bunch shown in Figure \ref{fig:bunch}) have roughly the same phases.
The bunch requires a longitude size smaller than
the halfwavelength, which is $\sim 10$ cm for GHz wave.

Inside a bunch, the total emission intensity is the summation of that of single charges with coherently adding.
Consider a single charge that moves along a trajectory $\boldsymbol{r}_s(t)$.
The energy radiated per unit solid angle per unit frequency interval is interested in considering spectrum, which is given by \citep{1979rpa..book.....R,1998clel.book.....J}
\be
\begin{aligned}
\frac{{\rm d}^2W}{{\rm d}\omega {\rm d}\Omega}&=\frac{e^2\omega^2}{4\pi^2c}\\
&\times\left|\int^{+\infty}_{-\infty}\sum_s^{N_s}-\boldsymbol{\beta}_{s\perp}\exp{[i\omega(t-\boldsymbol{n}\cdot\boldsymbol{r}_s/c)]}{\rm d}t\right|^2,
\label{eq:dW/domegadOmega}
\end{aligned}
\ee
where $\boldsymbol{\beta}_{s\perp}$ is the component of $\beta_s$ in the plane that is perpendicular to the LOS:
\be
\boldsymbol{\beta}_{s\perp}=-\boldsymbol{n}\times(\boldsymbol{n}\times\boldsymbol{\beta}_{s}).
\label{eq:bunchsum}
\ee
$\boldsymbol{\epsilon}_1$ is the unit vector pointing to the center of the instantaneous circle, $\boldsymbol{\epsilon}_2=\boldsymbol{n}\times\boldsymbol{\epsilon}_1$ is defined.
If there is more than one charged particle, one can use subscripts $(i,\,j,\,k)$ to describe any charge in the three-dimensional bunch.
The subscripts contain all the information about the location.
We assume that bunches are uniformly distributed in all directions of ($\hat{\boldsymbol{r}},\,\hat{\boldsymbol{\theta}},\hat{\boldsymbol{\varphi}}$).
Equation (\ref{eq:bunchsum}) can be written as
\begin{equation}
\begin{aligned}
\frac{{\rm d}^2W}{{\rm d}\omega {\rm d}\Omega}&=\frac{e^2 \omega^2}{4 \pi^2 c} \\
& \times\left|\int_{-\infty}^{+\infty} \sum_i^{N_i} \sum_j^{N_j} \sum_k^{N_k}-\boldsymbol{\beta}_{i j k,\perp} \exp\left[i \omega\left(t-\vec{n} \cdot \vec{r}_{i j k}(t) / c\right)\right] {\rm d}t\right|^2,
\end{aligned}
\end{equation}
where $N_{\mathrm{e}}=N_iN_jN_k$.
The emission power from charges with the same phase is $N_{\mathrm{e}}^2$ enhanced.
Detailed calculations are referred to Appendix \ref{app:radiation}.

The bunches thus act like single macro charges in some respects.
Define a critical frequency of curvature radiation $\omega_{\mathrm{c}}=3c\gamma^3/(2\rho)$, where $\rho$ is the curvature radius.
Spectra of curvature radiation from a single charge can be described as a power law for $\omega\ll\omega_{\mathrm{c}}$ and exponentially drop when $\omega\gg\omega_{\mathrm{c}}$.
The spectrum and polarization properties of a bunch with a longitude size smaller than the half-wavelength and half-opening angle $\varphi_t\lesssim1/\gamma$, are similar to the case of a single charge.
In general, the spectra can be characterized as multisegmented broken power laws by considering a variety of geometric conditions of bunch \citep{2018ApJ...868...31Y,2022ApJ...927..105W}.

\subsubsection{Bunch Structure}\label{sec2.1.1}

There could be a lot of bunches contributing to
instantaneous radiation in a moving bulk and their emissions are added incoherently.
The spread angle of relativistic charge is
\begin{equation}
\theta_c(\omega) \simeq\left\{
\begin{aligned}
&\frac{1}{\gamma}\left(\frac{2 \omega_{\mathrm{c}}}{\omega}\right)^{1 / 3}=\left(\frac{3 c}{\omega \rho}\right)^{1 / 3}&, \,\omega \ll \omega_{\mathrm{c}}\\
&\frac{1}{\gamma}\left(\frac{2 \omega_{\mathrm{c}}}{3 \omega}\right)^{1 / 2}&, \,\omega \gg \omega_{\mathrm{c}}
\label{eq:spreadangle}
\end{aligned}
.\right.
\end{equation}
For $\omega\sim\omega_{\mathrm{c}}$, bunches within a layer of $\rho/\gamma$ can contribute to the observed luminosity at an epoch.
The number of that contributing charges is estimated as $N_{\mathrm{b}}\sim2\rho\nu/(c\gamma)\simeq10^4\rho_7\nu_9\gamma_2^{-1}$.

We assume that the number density in a bunch has $\partial n_{\mathrm{e}}/\partial t=0$.
The case of fluctuating bunch density can give rise to a suppressed spectrum, and there is a quasi-white noise in a wider band in the frequency domain \citep{2023MNRAS.522.4907Y}.
The fluctuating density may cause the integral of Equation (\ref{eq:specA4}) in Appendix \ref{app:radiation} to not converge.

Different bunches may have different bulk Lorentz factors.
Note that emissions from relativistic charges may have a characteristic frequency determined by the Lorentz factor.
The scattering of the Lorentz factor from different emitting units likely enables the characteristic frequency to be polychromatic.
Therefore, non-monoenergetic distributed charges give rise to a wide spectrum.
However, for the bunch case discussed in Section \ref{sec2}, the layer contributing to the observed luminosity simultaneously is $\sim\rho/\gamma$ much smaller than the curvature radius.
Emission power in the layers is thus roughly the same.
The Lorentz factors between different bunch layers can be regarded as a constant.

The frequency structure of a burst is essentially the Fourier transform of the spatial structure of the radiating charge density \citep{2018MNRAS.481.2946K}.
Therefore, a relatively narrow spectrum may be produced via the emission from a quasi-periodically structured bulk of bunches.
The possible mechanisms that generate such quasi-periodically structure are discussed in Section \ref{sec3}.

\subsubsection{Spectrum of the Bunch Structure}\label{sec2.1.2}

\begin{figure*}
\resizebox{\hsize}{!}{\includegraphics{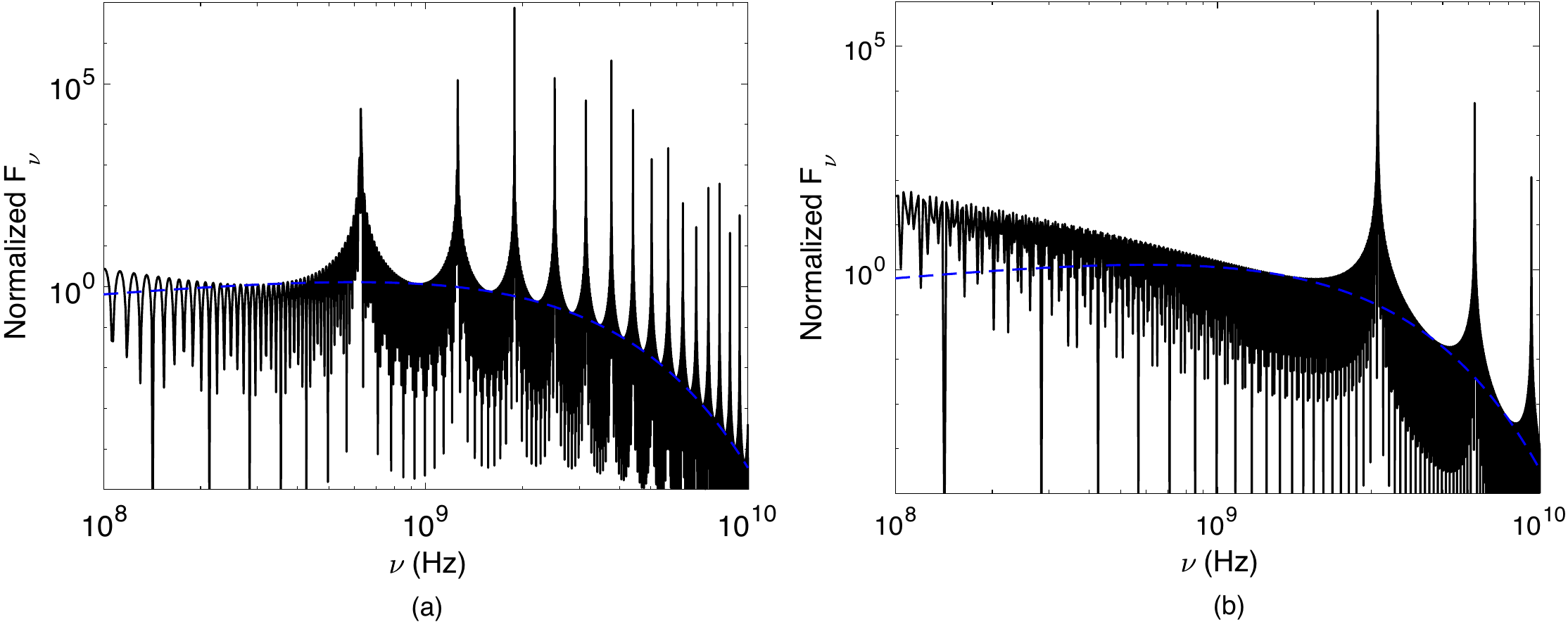}}
\caption{Normalized spectra of the quasi-periodically structured bulk of bunches: (a) $\omega_M=10^8$ Hz; (b) $\omega_M=5\times10^8$ Hz. The black solid line is the spectrum due to the quasi-periodic structure. The blue dashed line is the spectrum of a uniformly distributed bulk of bunches.}
\label{fig:bmspec}
\end{figure*}

We focus on the radiation properties due to the quasi-periodical structure.
The charged bunch distribution is thought to be quasi-period in the time domain, which may be attributed to some quasi-periodic distribution in space between bunches.
Each emitting bunch has the same $E(t)$ but with different arrival times.
The total received electric field from the multiple bunches simultaneously is given by 
\be
E(t)=\sum_n^{N_{\mathrm{b}}}E_n(t-t_n),
\ee
and its time-shifting property of the Fourier transform is
\be
E(\omega)=\sum_n^{N_{\mathrm{b}}}E_n(\omega)\exp(i\omega t_n).
\ee
A quasi-periodically distributed bulk of bunches means that $\omega t_n=n\omega/\omega_M+\delta\phi_n$, where $1/\omega_M$ denotes the period of the bunch distribution, and $\phi_n$ is the relative random phase.

A strictly periodically distributed bulk means that the relative random phase is zero strictly.
The energy per unit time per unit solid angle is \citep{2023ApJ...956...67Y}
\be
\frac{{\rm d}^2W}{{\rm d}\omega {\rm d}\Omega}=c^2\mathcal{R}^2\left|E_n(\omega)\right|^2\frac{\sin^2[N_{\mathrm{b}}\omega/(2\omega_M)]}{\sin^2[\omega/(2\omega_M)]},
\label{eq:quasiperiod}
\ee
where $\mathcal{R}$ is the distance from the emitting source to the observer.
The coherence properties of the radiation by
the multiple bulks of bunches are determined by $\omega_M$.
The radiation is coherently enhanced when $\omega$ is $2\pi\omega_M$ of integer multiples so that the energy is radiated into multiples of $2\pi\omega_M$ with narrow bandwidths.
Even $\phi_n$ is not zero, the Monte Carlo simulation shows that the coherent radiation energy is still radiated into multiples of $2\pi\omega_M$ with narrow bandwidths \citep{2023ApJ...956...67Y}.

Figure \ref{fig:bmspec} shows the simulation of radiation spectra from quasi-periodically distributed bulk of bunches with $\omega_M=10^8$ Hz and $\omega_M=5\times10^8$ Hz.
We assume that $N_{\mathrm{b}}=10^4$, $\gamma=10^2$, $\delta\phi_n=0$ and $\rho=10^7$ cm.
Since the flux drops to a small number rapidly when both $1/\gamma<\varphi'$ and $1/\gamma<\chi'$, we fix $\chi'=10^{-3}$ in the simulations for simplicity.
Spectra from uniformly distributed bulk of bunches are plotted as comparisons.
By assuming there is an observation threshold exceeding the peak flux of the pure curvature radiation, the spectrum is characterized by multiple narrow band emissions with a bandwidth smaller than $2\pi\omega_M$.
As $\omega_M$ becomes larger, the number of spikes is smaller.
Even though there are multi-spikes in the spectra, there is a main maximum at near $\nu=\nu_{\rm c}$ ordered larger than other spikes, so that multi-narrow-band emissions are difficult to detect from an ultra-wide-band receiver.

\subsection{Perturbation on Curved Trajectories}\label{sec2.2}

We consider two general cases that the perturbations exist both parallel ($E_{1,\|}$) and perpendicular ($E_{1,\perp}$) to the local $\boldsymbol{B}$-field independently.
The perturbations can produce extra parallel $c\dot{\beta}_\|$ and perpendicular accelerations $c\dot{\beta}_{\perp}$ to the local $\boldsymbol{B}$-field, as shown in Figure \ref{fig:bunch}.
The perturbation electric fields are much smaller than the $E_{\|}$, which is $\sim10^7$ esu, sustaining a constant Lorentz factor \citep{2019ApJ...876L..15W}.
Affected by these accelerations, electrons not only emit curvature radiation, but also emissions caused by the perturbations on the curved trajectory.
As a general discussion, we do not care about what is the perturbation in this Section but require that the velocity change slightly, i.e., $\delta \beta\ll\beta$.

\subsubsection{Parallel Acceleration}\label{sec2.2.1}

\begin{figure}
\resizebox{\hsize}{!}{\includegraphics{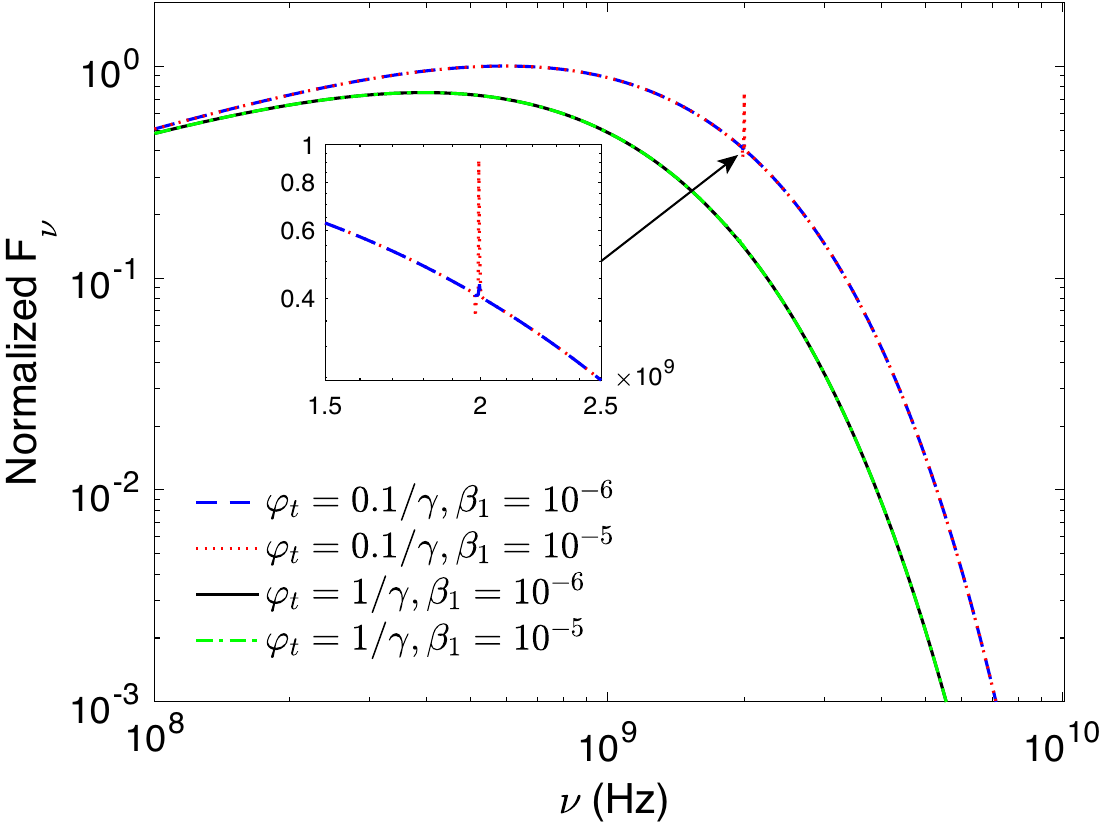}}
\caption{Normalized spectra of the parallel perturbation cases at $\varphi=0$: $\varphi_t=0.1/\gamma$, $\beta_{1}=10^{-6}$ (blue dashed line); $\varphi_t=0.1/\gamma$, $\beta_1=10^{-5}$ (red dotted line); $\varphi_t=1/\gamma$, $\beta_1=10^{-6}$ (black solid line); and $\varphi_t=1/\gamma$, $\beta_1=10^{-5}$ (green dashed-dotted line).}
\label{fig:specpa}
\end{figure}

The emission power by the acceleration parallel to the velocity, i.e., longitudinal acceleration, is given by \citep{1979rpa..book.....R}
\be
P=\frac{2e^2}{3m_{\mathrm{e}}^2c^3}\left(\frac{{\rm d}p}{{\rm d}t}\right)^2.
\label{eq:Ppara}
\ee
According to Equation (\ref{eq:Ppara}), if the ratio of two longitudinal forces is $N$, the ratio of their emission power to $N^2$, so that we can ignore the radiation damping caused by $E_{1,\|}$.
The energy loss by radiation is mainly determined by the coherent curvature radiation.
The balance between the $E_\|$ and radiation damping is still satisfied approximately, leading to a constant Lorentz factor.
Thus, the velocity change caused by $E_{1,\|}$ of a single charge can be described as
\be
E_{1,\|}e{\rm d}t\simeq m_{\mathrm{e}}\gamma_s^3{\rm d}v_s.
\label{eq:E1par}
\ee
The velocity change is so slight that the partial integration of Equation (\ref{eq:specA4}) converges, thus Equation (\ref{eq:specA5}) can be obtained from Equation (\ref{eq:specA4}).

\subsubsection{Spectrum of Parallel Acceleration}\label{sec2.2.2}

The dimensionless velocity of a single charge reads $\beta_s=\beta_{s,0}+\beta_{1,\|}$, where $\beta_{s,0}$ is the mean dimensionless velocity and $\beta_{1,\|}$ is its perturbation parallel to the field lines, so that the emitting electric field can be calculated by the curvature radiation and perturbation-induced emission independently.
Let us consider a charge with a dentifier of $i,j,k$ in a bunch.
The polarization components of the parallel perturbation-induced wave are calculated as
\be
\begin{aligned}
\tilde{A}_{ijk,\|}&\simeq\int^{+\infty}_{-\infty}\beta_{1,\|}(t)\sin\left(\frac{ct}{\rho}+\chi_{ij}\right)\\
&\times\exp\left(i \frac{\omega}{2}\left[\left(\frac{1}{\gamma^{2}}+\varphi_k^{2}+\chi_{ij}^{2}\right) t+\frac{c^{2} t^{3}}{3 \rho^{2}}+\frac{c t^{2}}{\rho} \chi_{ij}\right]\right){\rm d}t,\\
\tilde{A}_{ijk,\perp}&\simeq\int^{+\infty}_{-\infty}\beta_{1,\|}(t)\sin\varphi_k\cos\left(\frac{ct}{\rho}+\chi_{ij}\right)\\
&\times\exp\left(i \frac{\omega}{2}\left[\left(\frac{1}{\gamma^{2}}+\varphi_k^{2}+\chi_{ij}^{2}\right) t+\frac{c^{2} t^{3}}{3 \rho^{2}}+\frac{c t^{2}}{\rho} \chi_{ij}\right]\right){\rm d}t.\\
\end{aligned}
\label{eq:AmplitudePar}
\ee

A complex spectrum can be caused by a complexly varying $\beta_{1,\|}(t)$.
A simple case is that the integrals of Equation (\ref{eq:AmplitudePar}) can be described by the modified Bessel function $K_\nu(\xi)$ if $\beta_1$ is a constant.
The function is approximated to be power-law-like for $\xi\ll1$ and exponential-like for $\xi\gg1$ when $\nu\neq0$.
Here, it is worth noting that $K_\nu(\xi=0)\rightarrow{+\infty}$ when $\xi=0$, exhibiting a very narrow spectrum.

In particular, if a waveform $\boldsymbol{E}(t)$ is quasi-sinusoid periodic, the spectrum would be
quasi-monochromatic, equivalently to narrow-band.
The quasi-sinusoid periodic waveform can be produced by the intrinsic charge radiation if the charge is
under a periodic acceleration during its radiation beam pointing to the observer.
For a perturbation $E_{1.\|}$ which is monochromatic oscillating with time of interest, according to Equation (\ref{eq:E1par}) in Appendix \ref{app:radiation}, the velocity is also monochromatic oscillating.
Therefore, we introduce a velocity form with monochromatic oscillating in the lab frame, i.e., $\beta_{1,\|}(t)=\beta_1\exp(-i\omega_at)$.
The two parameters introduced in the curvature radiation in Equation (\ref{eq:uxi}) are replaced by
\be
\begin{aligned}
&u_a=\frac{c t}{\rho}\left(\frac{1}{\gamma^{2}}+\varphi_{k}^{2}+\chi_{ij}^2-\frac{2\omega_a}{\omega}\right)^{-1 / 2}, \\
&\xi_a=\frac{\omega \rho}{3 c}\left(\frac{1}{\gamma^{2}}+\varphi_{k}^{2}+\chi_{ij}^2-\frac{2\omega_a}{\omega}\right)^{3 / 2}.
\end{aligned}
\label{eq:newuxi}
\ee
The radiation can be narrow-band when $\gamma$ is a constant and the center frequency is
\be
\omega=\frac{2\omega_a\gamma^2}{1+\gamma^2(\chi_{ij}^2+\varphi_k^2)}.
\label{eq:narrowband}
\ee

For the bunching emission, charges are located with different $\chi$ and $\varphi$ which results in the bandwidth getting wider.
Now let's consider the change of the $E_\|$ is much smaller than $l\partial E_\|/\partial r \ll E_\|$, where $l$ is the longitude bunch size.
The charges in the bunch are then monoenergic.
However, for an ultra-relativistic particle, a small change of velocity can create a violent change of Lorentz factor, i.e., $\delta\gamma\simeq\gamma^3\beta\delta\beta_\|$.
To keep electrons being monoenergic, it is required that the amplitude $\beta_1\ll1/\gamma^2$.

Following the Appendix \ref{app:radiation}, one can obtain the spectrum of bunching charges.
Adopting $\gamma=100$, $\rho=10^7$ cm, $\varphi=0$ and $\omega_a=2\pi\times10^5\,\rm Hz$,
we simulate the spectra in four cases: $\varphi_t=0.1/\gamma$, $\beta_{1}=10^{-6}$; $\varphi_t=0.1/\gamma$, $\beta_1=10^{-5}$; $\varphi_t=1/\gamma$, $\beta_1=10^{-6}$; and $\varphi_t=1/\gamma$, $\beta_1=10^{-5}$, shown as Figure \ref{fig:specpa}.
The spectra are normalized to the $F_\nu$ of curvature radiation with $\varphi_t=10^{-3}$ at $\omega=\omega_{\mathrm{c}}$.
For $\varphi_t=0.1/\gamma$, the spectra show spikes at $2\nu_a\gamma^2$.
The amplitude of spikes decreases as $\beta_1$ decreases.
For $\varphi_t=1/\gamma$, the scattering of $\varphi'$ leads the spikes to become wider and shorter.
The small amplitude $\beta_1$ lets the spike disappear, and the spectrum is then determined by the curvature radiation.
As $\varphi$ becomes larger, more different trajectories may have more different azimuths so that the coherence of emission decreases.
The spectrum has a flat component at $\omega<\omega_{\mathrm{c}}$ when $3c/\rho<\omega_{\mathrm{c}}(\chi'^2+\varphi'^2)^{3/2}$ \citep{2018ApJ...868...31Y}.

\subsubsection{Perpendicular Acceleration}\label{sec2.2.3}

For comparable parallel and perpendicular forces ${\mathrm d}p/{\mathrm d}t$, the radiation power from the parallel component is of order $1/\gamma^2$ compared to that from the perpendicular component.
The energy loss by radiation is also mainly determined by the coherent curvature radiation.
Similar to the parallel scenario, the velocity change caused by $E_{1,\perp}$ of a single charge can be described as
\be
E_{1,\perp}e{\mathrm d}t\simeq m_{\mathrm{e}}\gamma_sc{\mathrm d}\beta_{1,\perp},
\ee
when $\beta_{1,\perp}\ll\beta_{\|}$.
The velocity change is also slight, therefore, one can use Equation (\ref{eq:specA5}) to calculate the spectrum (see Appendix \ref{app:radiation}).

\subsubsection{Spectrum of Perpendicular Acceleration}\label{sec2.2.4}

\begin{figure}
\resizebox{\hsize}{!}{\includegraphics{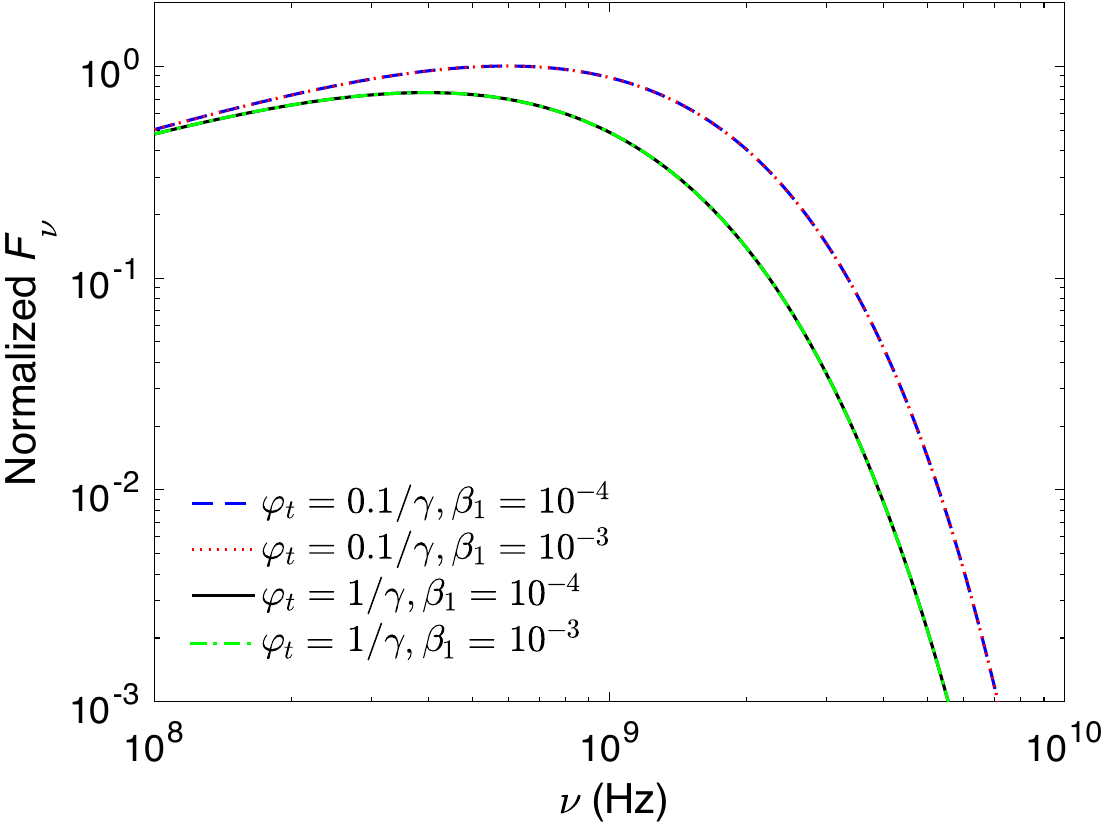}}
\caption{Normalized spectra of the perpendicular perturbation cases at $\varphi=0$: $\varphi_t=0.1/\gamma$, $\beta_{1}=10^{-4}$ (blue dashed line); $\varphi_t=0.1/\gamma$, $\beta_1=10^{-3}$ (red dotted line); $\varphi_t=1/\gamma$, $\beta_1=10^{-4}$ (black solid line); and $\varphi_t=1/\gamma$, $\beta_1=10^{-3}$ (green dashed-dotted line).
}
\label{fig:specpe}
\end{figure}

The polarization components of the perpendicular perturbation-induced wave of a charge with an identifier of $ijk$ are calculated as
\be
\begin{aligned}
\tilde{A}_{ijk,\|}&\simeq\int^{+\infty}_{-\infty}\frac{1}{2}\beta_{1,\perp}(t)\left[\varphi_k^2-(\chi_{ij}+ct/\rho)^2\right]\\
&\times\exp\left(i \frac{\omega}{2}\left[\left(\frac{1}{\gamma^{2}}+\varphi_k^{2}+\chi_{ij}^{2}\right) t+\frac{c^{2} t^{3}}{3 \rho^{2}}+\frac{c t^{2}}{\rho} \chi_{ij}\right]\right){\rm d}t,\\
\tilde{A}_{ijk,\perp}&\simeq\int^{+\infty}_{-\infty}\beta_{1,\perp}(t)\varphi_k (\chi_{ij}+ct/\rho) \\
&\times\exp\left(i \frac{\omega}{2}\left[\left(\frac{1}{\gamma^{2}}+\varphi_k^{2}+\chi_{ij}^{2}\right) t+\frac{c^{2} t^{3}}{3 \rho^{2}}+\frac{c t^{2}}{\rho} \chi_{ij}\right]\right){\rm d}t.\\
\end{aligned}
\label{eq:AmplitudePer}
\ee
where $\chi_{ij}$ and $\varphi_k$ are small angles.
The amplitudes are much smaller than those of parallel perturbations with comparable $\beta_1$.
Similar to the parallel scenario, the total amplitudes of a bunch for the perpendicular perturbation case can be calculated by summation of Equation (\ref{eq:bunchA2}).

A narrow spectrum is expected when $\gamma$ is a constant with the scattering of $\chi$ and $\varphi$ is small.
The spike is at $\omega=2\gamma^2\omega_a$ resembling that of the parallel perturbation case.
A small change in velocity can also make a violent change in the Lorentz factor but with a noticeable difference from that of longitude motion.
Under the energy supply from the $E_\|$, charges have $\beta_\|\approx1$.
The change of Lorentz factor is $\delta\gamma\simeq\gamma^3\beta_\perp\delta\beta_\perp$ when $\beta_{1,\perp}\ll1$.
To keep electrons being monoenergic, the condition of $\beta_1\ll1/\gamma$ should be satisfied at least.

Following the calculation steps in the Appendix \ref{app:radiation}, we simulate the spectra in four cases: $\varphi_t=0.1/\gamma$, $\beta_{1}=10^{-4}$; $\varphi_t=0.1/\gamma$, $\beta_1=10^{-3}$; $\varphi_t=1/\gamma$, $\beta_1=10^{-4}$; and $\varphi_t=1/\gamma$, $\beta_1=10^{-3}$, shown as Figure \ref{fig:specpe}.
Here, we take $\gamma=100$, $\rho=10^7$ cm, $\varphi=0$ and $\omega_a=2\pi\times10^5\,\rm Hz$.
The spectra are normalized to the $F_\nu$ of curvature radiation with $\varphi_t=10^{-3}$ at $\omega=\omega_{\mathrm{c}}$.
The spikes for the four cases are very small so the spectra are dominated by curvature radiations.

\subsection{Polarization of the Intrinsic Radiation Mechanisms}\label{sec2.3}

\begin{figure}
\resizebox{\hsize}{!}{\includegraphics{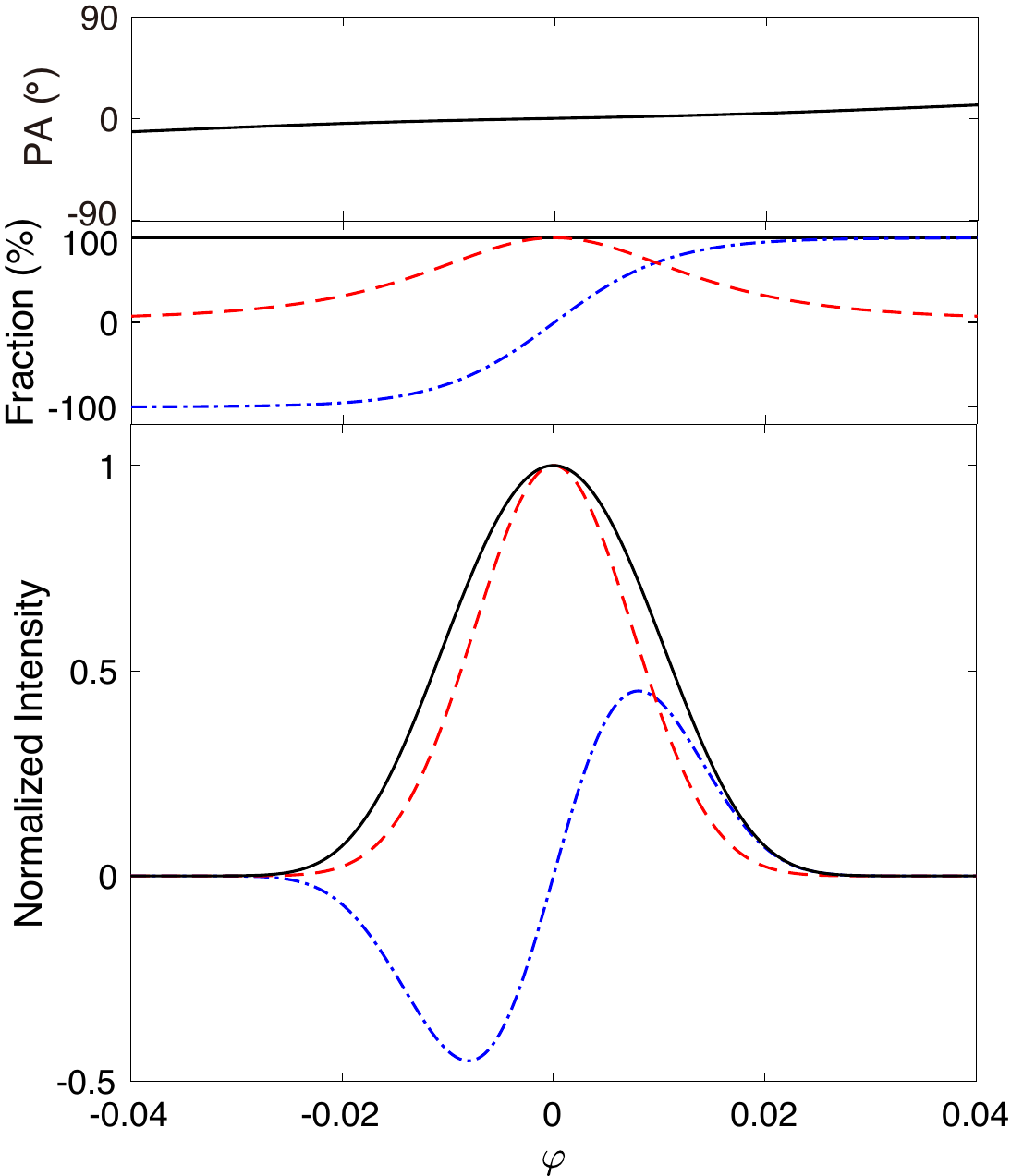}}
\caption{Simulated polarization profiles for the bunch with $\gamma=10^2$, $\varphi_t=10^{-2}$, $\rho=10^7$ cm. Top panel: the polarization position angle envelope. Middle panel: total polarization fraction (black solid line), linear polarization fraction (red dashed line), and circular polarization fraction (blue dashed-dotted line). Bottom: total intensity $I$ (black solid line), linear polarization $L$ (red dashed line), and circular polarization $V$ (blue dashed-dotted line) as functions of $\varphi$. All three functions are normalized to the value of $I$ at $\varphi=0$.
}
\label{fig:polarization}
\end{figure}

Four Stokes parameters can be used to describe the polarization properties of a quasi-monochromatic electromagnetic wave (see Appendix \ref{app:stokes}).
Based on Equation (\ref{eq:specA5}), the electric vectors of the emitting wave reflect the motion of charge in a plane normal to the LOS.
For instance, if the LOS is located within the trajectory plane of the single charge, the observer will see electrons moving in a straight line so that the emission is $100\%$ linearly polarized.
Elliptical polarization would be seen if the LOS is not confined to the trajectory plane.
We define that $\varphi < \theta_c$ as on-beam and $\varphi>\theta_c$ as off-beam.
For the bunch case, similar on-beam and off-beam cases can also be defined in which $\theta_c$ is replaced by $\varphi_t+\theta_c$ \citep{2022MNRAS.517.5080W}.
The general conclusion of the bunching curvature radiation is that high linear polarization appears for the on-beam geometry, whereas high circular polarization is present for the off-beam geometry.

We consider the polarization properties of the bunch structure case (Section \ref{sec2.1.2}), and the perturbation scenarios including the parallel and the perpendicular cases (Section \ref{sec2.2}).
For the quasi-periodically distributed bulk of bunches scenario, both $A_\|$ and $A_\perp$ multiply the same factor, so that the polarization properties are similar with a bulk of bunches, which bunches are {uniformly} distributed (e.g., \citealt{2022ApJ...927..105W}).
For the perturbation scenario, the perturbation on the curved trajectory can bring new polarization components, which generally obey the on(off)-beam geometry.
However, the intensity of the perturbation-induced components is too small then the polarization profiles are mainly determined by the bunching curvature radiation.
Consequently, the polarization properties for both the bunching mechanism and the perturbation scenarios can be described by the uniformly distributed bulk of bunches.

We take $\gamma=10^2$, $\varphi_t=10^{-2}$ and $\rho=10^7$ cm.
The beam angle at $\omega=\omega_{\rm c}$ is 0.02.
According to the calculations in Appendix \ref{app:stokes}, we simulate the $I$, $L$, $V$, and PA profiles, shown in Figure \ref{fig:polarization}.
These four parameters can completely describe all the Stokes parameters.
The total polarization fraction is $100\%$ because the electric vectors are coherently added.
The polarization position angle (PA) for the linear polarization exhibits some variations but within $30^\circ$.
The differential of $V$ is the largest at $\varphi=0$ and becomes smaller when $|\varphi|$ gets larger.
High circular polarization with flat fractions envelope would appear at off-beam geometry.
In this case, the flux is smaller but may still be up to the same magnitude as the peak of the profile (e.g., \citealt{2023ApJ...943...47L}).
The spectrum at the side of the beam becomes flatter due to the Doppler effect \citep{2021ApJ...907L..17Z}.
The sign change of $V$ would be seen when $\varphi=0$ is confined into the observation window \citep{2022MNRAS.517.5080W}.

\section{Trigger of Bunches}\label{sec3}

The trigger mechanism plays an important role in producing bunching charges and creating the $E_\|$.
Intriguingly, a giant glitch was measured $3.1\pm2.5$ days before FRB 20200428A associated with a quasi-periodic oscillations (QPOs) of $\sim40$ Hz in the X-ray burst from the magnetar SGR J1935+2154 \citep{2020ApJ...898L..29M,2021NatAs...5..378L,2021NatAs...5..372R,2021NatAs...5..401T,2022arXiv221103246G,2022ApJ...931...56L}.
Another glitch event was found on 5 October 2020 and the magnetar emitted three FRB-like radio bursts in subsequent days \citep{2023NatAs...7..339Y,2023SciA....9F6198Z}.
Both the glitches and QPOs are most likely the smoking guns of starquakes.
Therefore, magnetar quakes are considered a promising trigger model of FRB in the following.

As discussed in Section \ref{sec2.1.2}, we are particularly interested in the mechanisms that can generate quasi-periodically distributed bunches.
Two possible scenarios are discussed independently in Section \ref{sec3.1.1} and Section \ref{sec3.1.2}.
Motivated by the fact that Alfv\'en waves associated with the quake process have been proposed by lots of models (e.g., \citealt{2018ApJ...852..140W,2020MNRAS.494.2385K,2021ApJ...919...89Y,2022ApJ...933..174Y}),
we discuss the Alfv\'en waves as the perturbations in Section \ref{sec2.2}.

\subsection{Structured Bunch Generation}\label{sec3.1}

\subsubsection{Pair Cascades from ``Gap''}\label{sec3.1.1}

A magnetized neutron star's rotation can create a strong electric field that extracts electrons from the stellar surface and accelerates them to high speeds \citep{1975ApJ...196...51R}.
The induced electric field region lacks adequate plasma, i.e., charge-starvation, referred to as a ``gap''.
The extracted electrons move along the curved magnetic field lines near the surface, emitting gamma rays absorbed by the pulsar's magnetic field, resulting in electron-positron pairs.
Such charged particles then emit photons that exceed the threshold for effective pair creation, resulting in pair cascades that populate the magnetosphere with secondary pairs.
The electric field in the charge-starvation region is screened once the pair plasma is much enough and the plasma generation stops \citep{2005ApJ...631..456L}.
When the plasma leaves the region, a gap with almost no particles inside is formed forms again, and the vacuum electric field is no longer screened.

The numerical simulations show that such pair creation is quasi-periodic \citep{2010MNRAS.408.2092T,2022ApJ...933L..37T}, leading to quasi-periodically distributed bunches.
The periodicity of cascades is $\sim 3h_{\rm gap}/c$, where $h_{\rm gap}$ is the height of the gap.
The gap's height is written as \citep{2015ApJ...810..144T}
\begin{equation}
h_{\rm gap}=1.1 \times 10^4\rho_{7}^{2 / 7} P_0^{3 / 7} B_{12}^{-4 / 7}\left|\cos \alpha\right|^{-3 / 7}\,\mathrm{cm},
\end{equation}
where $P$ is the spin period and $\alpha$ is the inclination angle of magnetic axis.
For a typical radio pulsar, we take $B=10^{12}$ G and $P=1$ s.
The corresponding $\omega_M$ is $10^7$ Hz.
According to Equation (\ref{eq:quasiperiod}), radiation energy is emitted into multiples of $\sim10$ MHz narrow bandwidths, which is consistent with the narrow-band drifting structure of PSR B0950+08 \citep{2022A&A...658A.143B}.

In contrast to the continuous sparking in the polar cap region of normal pulsars, FRB models prefer a sudden, violent sparking process.
Starquake as a promising trigger mechanism can occur when the pressure induced by the internal magnetic field exceeds a threshold
stress.
The elastic and magnetic energy released from the crust into the magnetosphere are much higher than the spin-down energy for magnetars.
Crust shear and motion may create hills on the surface.
The height of the gap becomes smaller and the electric field is enhanced due to the so-called point-discharging.
Quasi-periodically distributed bunches with $\omega_M\sim10^8$ may be formed from such shorter gaps.
Observational consequences of small hills on stellar surfaces could also be known by studying regular pulsars \citep[e.g.,][]{2023arXiv230807691W}.

Note that the sparking process to generate FRB-emitting bunches is powered by magnetic energy rather than spin-down energy.
The stellar oscillations during starquakes can provide
additional voltage, making the star active via sparking \citep{2015ApJ...799..152L}.
The oscillation-driven sparking process may generate much more charges than the spin-down powered sparking of a normal rotation-powered pulsar.

\subsubsection{Two-stream Instability}\label{sec3.1.2}

Alternatively, another possibility to form charged bunches is the two-stream instability between electron-positron pairs.
The growth rate of two-stream instability is discussed by invoking linear waves.
Following some previous works \citep[e.g.,][]{1987ApJ...320..333U,2002MNRAS.337..422G,2023MNRAS.522.4907Y}, we consider that there are two plasma components denoted by ``1'' and ``2'' with a relative motion along the magnetic field line.
The Lorentz factor and the particle number density are $\gamma_j$ and $n_j$ with $j=1,2$. 
Each plasma component is assumed to be cold in the rest frame of each component.
Then the dispersion relation of the plasma could be written as \citep{1998MNRAS.301...59A}
\begin{align}
1-\frac{\omega_{{\rm p},j=1}^2}{\gamma_{j=1}^3(\omega-\beta_{j=1} kc)^2}-\frac{\omega_{{\rm p},j=2}^2}{\gamma_{j=2}^3(\omega-\beta_{j=2} kc)^2}=0,
\end{align}
where $\omega_{{\rm p},j}$ is the plasma frequency of the component $j$.
For the resonant reactive instability as the bunching mechanism, the growth rate and the characteristic frequency of the Langmuir wave are given by \citep[e.g.,][]{1987ApJ...320..333U,2002MNRAS.337..422G,2023MNRAS.522.4907Y}
\begin{align}
&\Gamma\sim\gamma_{j=1}^{-1}\gamma_{j=2}^{-1}\left(\frac{n_{j=1}}{n_{j=2}}\right)^{1/3}\omega_{\rm L},\\
&\omega_{\rm L}\sim2\gamma_{j=2}^{1/2}\omega_{{\rm p},j=2}.
\end{align}
The linear amplitude of the Langmuir wave depends on the gain $G\sim\Gamma r/c$. 
According to \citet{2020MNRAS.497.3953R}, a threshold gain $G_{\rm th}$ indicating the breakdown of the linear regime is about $G_{\rm th}\sim$ a few. The sufficient amplification would drive the plasma beyond the linear regime once $G\sim G_{\rm th}$, and the bunch formation rate is $\Gamma(G\sim G_{\rm th})$.
The bunch separation corresponds to the wavelength of the Langmuir wave for $G\sim G_{\rm th}$, i.e.,
\begin{align}
l_{\rm sep}\sim\frac{\pi c}{\gamma_{j=2}^{1/2}\omega_{{\rm p},j=2}}.
\end{align}
A narrow-band spectrum with $\Delta\nu\sim100$ MHz would be created if $\gamma_{j=2}^{1/2}\omega_{{\rm p},j=2}\sim10^8\,\rm Hz$.

Generally, the charged bunches would form and disperse dynamically due to the instability, charge repulsion, velocity dispersion, etc \citep[e.g.,][]{2021ApJ...918L..11L,2023MNRAS.522.4907Y}. The spectrum of the fluctuating bunches depends on the bunch formation rate $\lambda_B$, lifetime $\tau_B$, and bunch Lorentz factor $\gamma$, as proposed by \citet{2023MNRAS.522.4907Y}.
The coherent spectrum by a fluctuating bunch is suppressed by a factor of $(\lambda_B\tau_B)^2$ compared with that of a persistent bunch, and there is a quasi-white noise in a wider band in the frequency domain. The observed narrow spectrum implies that, at the minimum, the quasi-white noise should not be dominated in the spectrum, in which case, the condition of $2\gamma^2\lambda_B\gtrsim\min(\omega_{\rm peak},2\gamma^2/\tau_B)$ would be required, where $\omega_{\rm peak}$ is the peak frequency of curvature radiation,
see \citet{2023MNRAS.522.4907Y} for a detailed discussion.

\subsection{Quake-induced Oscillation and Alfv\'en Wave}\label{sec3.2}

\begin{figure}
\resizebox{\hsize}{!}{\includegraphics{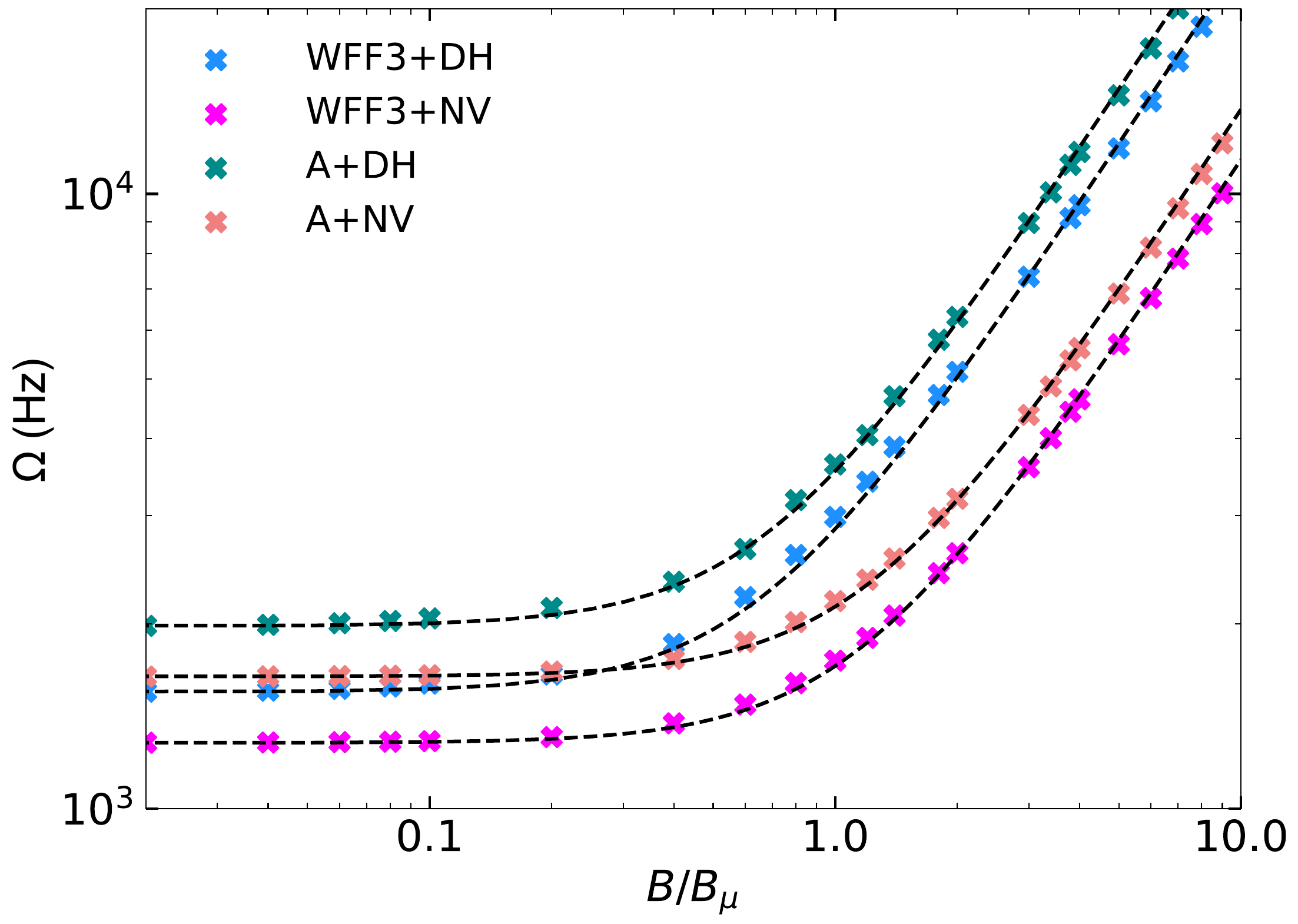}}
\caption{The frequencies of the second overtone $n=2$ and $\ell=2$ torsional modes as a function of the 
    normalized magnetic field ($B/B_{\mu}$). The NS mass is $M = 1.4\, M_{\odot}$. The dashed lines correspond to our fits using the empirical formula (\ref{eq: fit}) with different coefficient values. The fitting values are $2.4$, $0.78$, $2.2$, and $0.69$ for WFF3+ DH model, WFF3+ NV model, A+ DH model, and A+ NV model, respectively.}
\label{fig:f_NS}
\end{figure}

An Alfv\'en wave packet can be launched from the stellar surface during a sudden crustal motion and starquake.
The wave vector is not exactly parallel to the field line so the Alfv\'en waves have a non-zero electric current along the magnetic field lines.
Starquakes as triggers of FRBs are always accompanied by Alfv\'en wave launching.
We investigate whether Alfv\'en waves as the perturbations discussed in Section \ref{sec2.2} can affect the spectrum.

The angular frequencies of the torsional modes $\Omega$ caused by a quake are calculated as follows (also see Appendix \ref{app:quake}).
We chose the equation of state (EOS A) \citep{Pandharipande:1971} 
and EOS WFF3 \citep{Wiringa:1988}, which describe the core of neutron stars, as well as consider two different proposed EOSs
for the crusts, including EOSs for NV \citep{Negele:1973} and DH \citep{Douchin:2001sv} models.
We match the various core EOS to two different EOSs for the crust.
The crust-core boundary for the NV and DH EOSs is defined at $\rho_m \approx 2.4 \times 10^{14}\, \rm g\,cm^{-3}$, and $\rho_m \approx 1.28 \times 10^{14}\, \rm g\,cm^{-3}$, respectively.

The frequencies of torsional modes are various with a strong magnetic field \citep{Messios:2001br}.
As already emphasized by \citet{Sotani:2006at}, the shift in the frequencies would be significant when the magnetic field exceeds $\sim 10 ^{15}$ G.
In the presence of magnetic fields, frequencies are shifted as,
\begin{equation}
{}_\ell f_n = {}_\ell f_n^{(0)} \left[ 1+ {}_\ell \alpha_n
\left(\frac{B}{B_{\mu}}\right)^2 \right]^{1/2}\,,
\label{eq: fit}
\end{equation}
where ${}_\ell {\alpha}_n$ is a coefficient depending on the structure of the star, ${}_\ell f_n^{(0)}$ is the frequency of the non-magnetized star, and $B_\mu \equiv (4 \pi\mu)^{1/2}$ is the typical magnetic field strength.

In Fig.~\ref{fig:f_NS}, we show the effects of the
magnetic field on the frequencies of the torsional modes. The magnetic field strength is normalized by $B_{\mu} = 4 \times 10^{15}$\,G. Different dashed lines in Fig.~\ref{fig:f_NS} are our fits to the
calculated numerical data with a high accuracy.
For $B > B_{\mu}$, we find that the frequencies follow a quadratic increase against the magnetic field, and tend to become less sensitive to the NS parameters.
The oscillation frequencies exceed $10^3$ Hz for $n=2$ and $\Omega\sim10^4$ Hz for typical magnetars.

Alfv\'en waves are launched from the surface at a frequency of $\Omega$, which is much faster than the angular frequency of the spin, and propagate along the magnetic field lines.
Modifying the Goldreich-Julian density can produce electric fields to accelerate magnetospheric charged particles.
The charged particles oscillate at the same frequency as the Alfv\'en waves in the comoving frame.
In the lab frame, the perturbation has $\omega_a=\Omega\gamma^2/\gamma^2_{\rm A}$ due to the Doppler effect, where $\gamma_{\rm A}=\sqrt{1+B^2/(4\pi\rho_m c^2)}$.
$\omega_a$ is much smaller than $10^8\Omega_4\gamma^2_2\,\rm{Hz}$ because of the strong magnetic field and low mass density in the magnetosphere.

Therefore, by considering that Alfv\'en waves play a role in the electric perturbations discussed in Section \ref{sec2.2}, the spectra have no spike from 100 MHz to 10 GHz, which is the observation bands for FRBs so far.
The spectral spike would appear at the observation bands when the perturbations are global oscillations, meaning that all positions in the magnetosphere oscillate simultaneously with $\Omega$ rather than a local Alfv\'en wave packet.

\section{Absorption Effects in Magnetosphere}\label{sec5}

In this Section, we investigate absorption effects during the wave propagating the magnetosphere.
The absorption effects discussed as possible origins of narrow-band emission are independent of the intrinsic mechanisms in Section \ref{sec2}.
To create narrow-band emission, there has to be a steep frequency cut-off on the spectrum due to the absorption conditions.
Absorption conditions should also change cooperatively with the emission region, that is, the cut-off appears at a lower frequency as the flux peak occurs at a lower frequency.
Curvature radiation spectra can be characterized as a power law for $\omega<\omega_{\mathrm{c}}$ and exponential function at $\omega>\omega_{\mathrm{c}}$ \citep{1998clel.book.....J}.
In the following discussion, we mainly consider two absorption mechanisms including Landau damping and self-absorption of curvature radiation.

\subsection{Wave Modes}\label{sec5.1}
We consider the propagation of electromagnetic waves in magnetospheric relativistic plasma.
The magnetosphere mainly consists of relativistic electron-positron pair plasma, which can affect the radiation created in the inner region.
We assume that the wavelength is much smaller than the scale lengths of the plasma and of
the magnetic field in which the plasma is immersed.
The scale lengths of the magnetic field can be estimated as $B/(\partial B/\partial r)\sim r$.
The plasma is consecutive and its number density is thought to be multiples of the Goldreich-Julian density \citep{1969ApJ...157..869G}, leading to the same scale lengths as that of the magnetic field.
Both the scale lengths of magnetic field and plasma are much larger than the wavelength $\lambda=30\nu^{-1}_9$ cm, even for outer magnetosphere $r\sim4.8\times10^9 P_0$ cm.

Whatever a neutron star or a magnetar, the magnetospheric plasma is highly magnetized.
It is useful to define the (non-relativistic) cyclotron frequency and plasma frequency:
\be
\omega_B=\frac{eB}{m_{\mathrm{e}}c}=1.78\times10^{19}B_{12}\,\mathrm{Hz},
\label{eq:omegaB}
\ee
\be
\omega_{\mathrm{p}}=\left(\frac{4\pi n_{\mathrm{p}}e^2}{m_{\mathrm{e}}}\right)^{1/2}=4.73\times10^{11}\mathcal{M}^{1/2}_3B^{1/2}_{12}P^{1/2}_0\,\mathrm{Hz}.
\label{eq:omegap}
\ee
where $n_{\mathrm{p}}=\mathcal{M} n_{\rm GJ}=\mathcal{M} B/(ceP)$, in which $n_{\rm GJ}$ is the Goldreich-Julian density, and $\mathcal{M}$ is a multiplicity factor.
In the polar-cap region of a pulsar, a field-aligned electric field may accelerate primary particles to ultra-relativistic speed.
The pair cascading may generate a lot of plasma with lower energy and a multiplicity factor of $\mathcal{M}\sim10^2-10^5$ \citep{1982ApJ...252..337D}.
The relationships that $\omega\ll\omega_B$ and $\omega_{\mathrm{p}}\ll\omega_B$ are satisfied at most regions inside the magnetosphere.

The vacuum polarization may occur at the magnetosphere for high energy photons and radio waves by the plasma effect in a strong magnetic field \citep{1971AnPhy..67..599A}.
At the region of $r\gtrsim10R$ where FRBs can be generated, in which $R$ is the stellar radius, the magnetic field strength $B<B_Q=m_{\mathrm{e}}^2c^3/(e\hbar)$ with assumption of $B_{\rm s}=10^{15}$ at stellar surface.
In the low-frequency limit $\hbar \omega\ll m_{\mathrm{e}}c^2$, the vacuum polarization coefficient can be calculated as $\delta_V=1.6\times10^{-4}(B/B_Q)^2=2.65\times10^{-8}B_{12}^2$ \citep{1997PhRvD..55.2449H,2007MNRAS.377.1095W}.
Therefore, the vacuum polarization can be neglected in the following discussion about the dispersion relationship in the magnetosphere.

Waves have two propagation modes: O-mode and X-mode.
In the highly magnetized relativistic case, the dispersion of X-mode photon is $n\approx1$.
O-mode photons have two branches in the dispersion relationship, which are called subluminous O-mode ($n>1$) and superluminous O-mode ($n<1$), shown as Figure \ref{fig:modes}.
The subluminous O-mode cuts off at $1/\cos\theta_B$, where $\theta_B$ is the angle between angle between $\boldsymbol{\hat{k}}$ and $\boldsymbol{\hat{B}}$.
Following the calculation shown in Appendix \ref{app:dispersion}, the dispersion relation of magnetospheric plasma for subluminous O-mode photons can be found as \citep{1986ApJ...302..120A,1988Ap&SS.146..205B}
\begin{equation}
\left(\omega^2-c^2 k_{\|}^2\right)\left(1-\sum_s \frac{\omega_{{\rm p},s}^2}{\omega^2} g_s\right)-c^2 k_{\perp}^2=0,
\label{eq:DR}
\end{equation}
with
\begin{equation}
g_s=\mathcal{P} \int_{-\infty}^{\infty} \frac{f\left(u\right){\rm d}u}{\gamma^3\left(1-n_{\|} \beta\right)^2}+i \pi\beta_n^2\gamma_n^3 \left[\frac{{\rm d}f}{{\rm d}u}\right]_{\beta=\beta_n},
\label{eq:g}
\end{equation}
where $\beta_n=n_{\|}^{-1}=\omega/(k_{\|}c)$, $\gamma_n=(1-\beta_n^2)^{-1/2}$ and $f(u)$ is the distribution function of plasma.

\subsection{Landau Damping}\label{sec5.2}

\begin{figure}
\resizebox{\hsize}{!}{\includegraphics{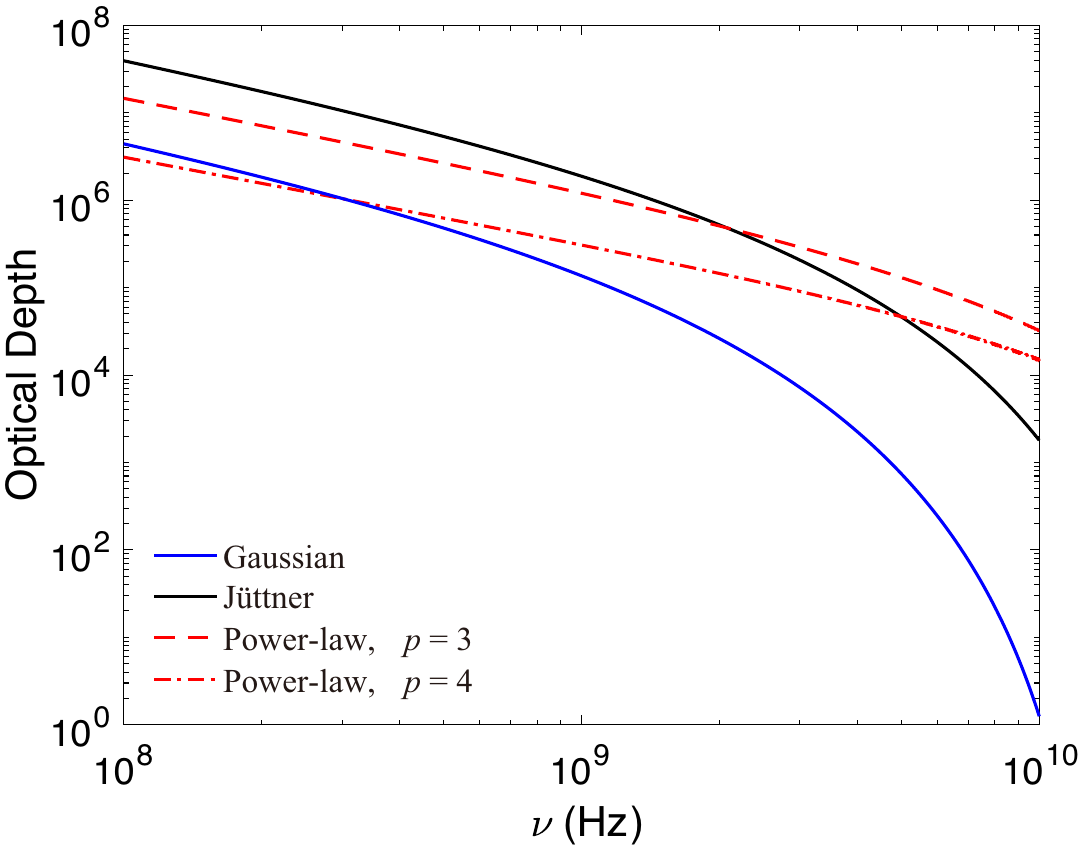}}
\caption{Optical depths for four distributions: J\"uttner with $\rho_T=0.1$ (black solid line), Gaussian with $u_m=10/\sqrt{2\pi}$ (blue solid line), power-law with $p=3$ (red dashed line) and power-law with $p=4$ (red dotted-dashed line).
}
\label{fig:subO}
\end{figure}

Wave forces can exist many collective waves in a beam and exchange energy between waves and particles.
If the distribution function is ${\mathrm d}f/{\mathrm d}v<0$ close to the phase speed of the wave, the number of particles with speeds $v < \omega/k$ will be larger than that of the particles having $v > \omega/k$.
Then the transmission of energy to the wave is then smaller than that absorbed from the wave, leading to a Landau damping of the wave.
According to the dispersion relationship given in Section \ref{sec5.1}, there is no linear damping of the X-mode since wave-particle resonance is absent.
The superluminous branch of the O-mode also has no Landau damping because the phase speed exceeds the speed of light.
Landau damping only occurs at the subluminous O-mode photons.

The Landau damping decrement can be calculated by the imaginary part of the dispersion relationship \citep{2003phpl.book.....B}.
We let $k=k_r+ik_i$, where $k_i\ll k_r\approx k$.
If a wave has a form of $E\propto\exp(-i\omega t+ikr)$, the imaginary $k_i$ can make the amplitude drop exponentially.
The imaginary part of Equation (\ref{eq:DR}) by considering a complex waveform of $k$ is given by
\be
k_i=\frac{n^2\omega_{\mathrm{p}}^2\cos^2\theta_B-\omega_{\mathrm{p}}^2}{2kc^2(1-g_r\omega_{\mathrm{p}}^2\cos^2\theta_B/\omega^2)}g_i,
\ee
where $g_r$ and $g_i$ denote the real and imaginary parts of Equation (\ref{eq:g}).
Consider emission at radius $r_e$ along a field line whose footprint intersects the crust at magnetic colatitude $\Theta_e(R/r_e)^{1/2}$ for $\Theta_e\ll1$, one can obtain the angle between $\boldsymbol{k}$ and $\boldsymbol{B}$ at each point:
\be
\theta_B=\frac{3}{8}\Theta_e\left(\frac{r}{r_e}\right)^{1/2}\left(1-\frac{r_e}{r}\right)^2.
\label{eq:thetae}
\ee
Consequently, the optical depth for waves emitted at $r_e$ to reach $r$ is
\be
\tau_{\rm LD}=\int^r_{r_e}2k_i dr.
\label{eq:opticaldepth}
\ee

Even the distribution function for the magnetospheric electrons and positrons is not known, a relativistic thermal distribution is one choice.
Since the vertical momentum drops to zero rapidly, for instance, a 1-D J\"uttner distribution is proposed here (e.g., \citealt{1999JPlPh..62..233M,2019JPlPh..85c9005R}):
\begin{equation}
f(u)=\frac{1}{2 K_1(\rho_T)} \exp(-\rho_T \gamma_{\rm p}),
\label{eq:juttner}
\end{equation}
where $\rho_T=m_{\mathrm{e}}c^2/(k_{\mathrm{B}}T)$, in which $T$ is the temperature of plasma and $k_{\mathrm{B}}$ is the Boltzmann constant.
Alternatively, another widely used distribution is a Gaussian distribution (e.g., \citealt{1983Ap.....19..426E,1998MNRAS.301...59A})
\begin{equation}
f(u)=\frac{1}{\sqrt{2\pi}u_m} \exp \left(-\frac{u^2}{2 u_m^2}\right),
\label{eq:Gaussian}
\end{equation}
where $u_m$ can be interpreted as the average $u$ over the distribution function.
Besides a thermal distribution, power law distributions $f(\gamma_{\rm p})=(p-1)\gamma_{\rm p}^{-p}$, where $p$ is the power law index ($p>2$), are representative choice to describe the plasma in the magnetosphere (e.g., \citealt{1973plas.book.....K}).

The optical depths as a function of wave frequency with four different distributions are shown in Figure \ref{fig:subO}.
Here, we assume that $r_e=10^7$ cm, $\mathcal{M}=10^3$, $\gamma_{\rm p}=10$ and $\Theta_e=0.05$.
For all considered distributions, optical depths decrease with frequencies which means that the Landau damping is more significant for lower-frequency subluminous O-mode photons.
The optical depths are much larger than 1 for the FRB emission band so that the subluminous O-mode photon can not escape from the magnetosphere.
Superluminous O-mode photons which frequency higher than $\omega_{\rm cut}=\omega_{\mathrm{p}}\gamma_{\rm p}^{-3/2}$ can propagate.
The bunching curvature radiation is exponentially dropped at $\omega>\omega_{\mathrm{c}}$.
If the superluminous O-mode cut-off frequency is close to $\omega_{\mathrm{c}}$, the spectrum of the emission would be narrow.
A simple condition could be given when a dipole field is assumed:
\be
r\simeq13.8R\mathcal{M}_3B_{s,15}P^{-1}_0\gamma^{-6}_2\gamma_{\rm p, 1.3}^{-3},
\ee
where the emission region is considered at closed field lines in which $\rho\sim r$.
The cut-off frequency $\omega_{\rm cut}\propto r^{-3/2}$ which evolves steeper than $\omega_{\mathrm{c}}\propto r^{-1}$.
The bandwidth is suggested to become wider as the burst frequency gets lower.

Coherent curvature radiation can propagate out of the magnetosphere via superluminous O-mode with a frequency cut-off, leading to narrow spectra likely.
However, the condition that the frequency cut-off is close to the characteristic frequency of curvature radiation could not be satisfied at all heights, leading to narrow spectra sometimes.
The waves can also propagate via X-mode photon which has no frequency cut-off.
The spectrum of X-mode photons ought to be broad.
The waves propagate via whether X-mode or O-mode is hard to know because the real magnetic configuration is complex.
Narrow spectra have been found in some FRBs, while some FRBs seem to have broad spectra, and there are also some not sure due to the limitation of bandwidth of the receiver \citep{2023ApJ...955..142Z}.

In principle, the spectrum of a radio magnetar would be narrow, if the emission is generated from $r> 10R$ propagating via O-mode.
However, radio pulses are rarely detected among magnetars.
For normal pulsars, the lower magnetic fields may induce lower number density so that the Landau damping is not as strong as in magnetars.

\subsection{Self-absorption by Curvature Radiation}\label{sec5.3}

\begin{figure}
\resizebox{\hsize}{!}{\includegraphics{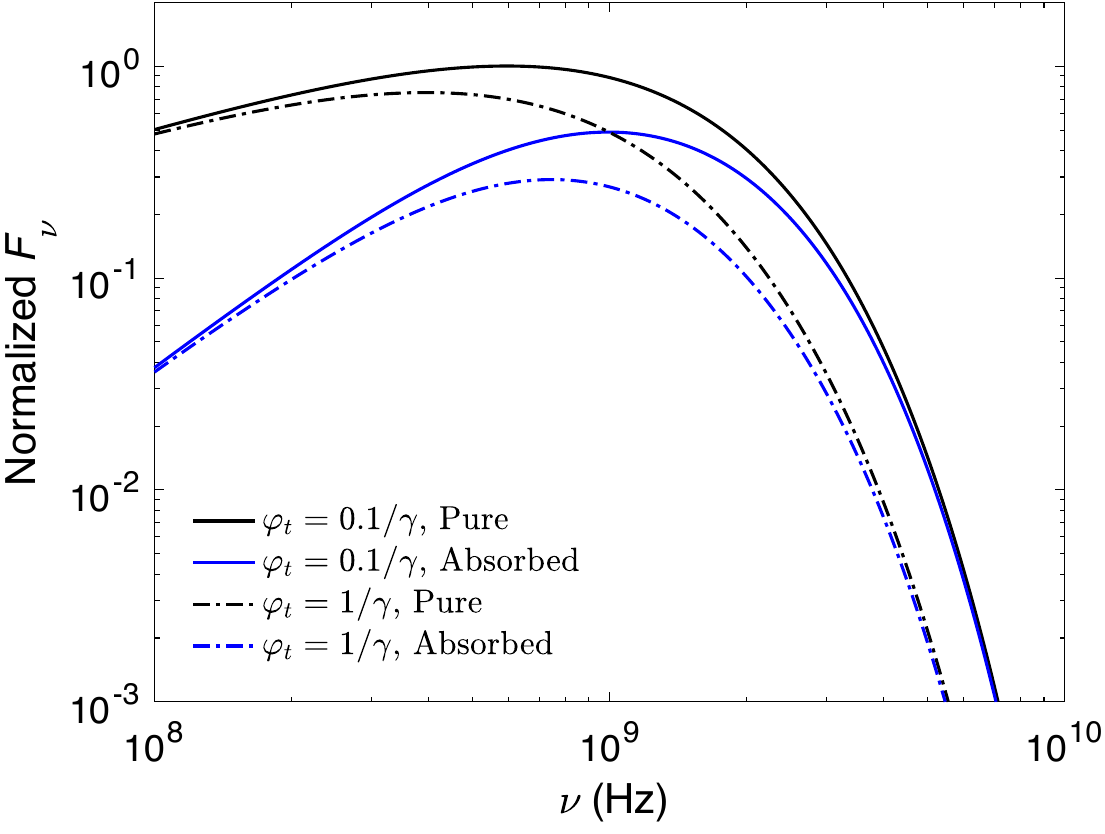}}
\caption{Normalized spectra of self-absorbed curvature radiation. Black lines denote purely bunching curvature radiations for $\varphi_t=0.1/\gamma$ (solid line) and $\varphi_t=1/\gamma$ (dashed-dotted line). Blue lines denote the radiation after curvature self-absorption for $\varphi_t=0.1/\gamma$ (solid line) and $\varphi_t=1/\gamma$ (dashed-dotted line).
}
\label{fig:SaCR}
\end{figure}

The self-absorption is also associated with the emission.
We assume that the momentum of the particle is still
along the magnetic field line even after having absorbed the photon.
The cross-section can be obtained through the Einstein coefficients when $h\nu\ll\gamma_sm_{\mathrm{e}}c^2$ (e.g., \citealt{2018A&A...617A..84L}):
\begin{equation}
\frac{{\rm d} \sigma_{\mathrm{c}}}{{\rm d} \Omega}=\frac{1}{2 m_{\mathrm{e}} \nu^2} \frac{\partial}{\partial \gamma_s}\left[j_s(\nu, \gamma_s, \varphi_s)\right],
\label{eq:SAcrosssection}
\end{equation}
where $\sigma_{\mathrm{c}}$ is the curvature cross section and $j_s$ is the emissivity.

There are special cases which the stimulated emission becomes larger than the true absorption.
As a result, the total cross section becomes negative, and there is a possible maser mechanism.
However, such maser action is ineffective when the magnetic energy density is much higher than that for the kinetic energy of charges \citep{1979AuJPh..32...49Z}.
The isotropic equivalent luminosity for the curvature radiation is $\sim10^{42}\,\rm erg\,s^{-1}$ if $N_{\mathrm{e}}=10^{22}$ at $r=10^7$ cm for $\gamma=10^2$ \citep{2017MNRAS.468.2726K}.
The number density in a bunch is estimated as $n_{\mathrm{e}}\sim10^{11}\,\rm cm^{-3}$ by \cite{2022ApJ...927..105W}, so that $n_{\mathrm{e}}\gamma m_{\mathrm{e}}c^2\ll B^2/(8\pi)$.
Thus, the maser process is ignored in the following discussion.

By inserting the curvature power per unit frequency (see e.g., \citealt{1998clel.book.....J}), the total  cross section integrated over angles is given by
\citep{2018A&A...613A..61G}
\begin{equation}
\sigma_{\mathrm{c}} \approx 2^{5/3}\sqrt{3} \Gamma(5 / 3)\frac{\pi e^2}{\gamma_s^3 m_{\mathrm{e}} c\nu}, \quad \nu \ll \nu_{\mathrm{c}},
\label{eq:sigmac}
\end{equation}
where $\Gamma(\nu)$ is the Gamma function.
As discussed in Section \ref{sec2}, the charges are approximately monoenergetic due to the rapid balance between $E_\|$ and the curvature damping.
The absorption optical depth is
\be
\tau_{\rm CR}\simeq\sigma_{\mathrm{c}}n_{\mathrm{e}}\frac{\rho}{\gamma_s}.
\ee
The absorption is more significant for lower frequencies.

Figure \ref{fig:SaCR} exhibits the comparison between purely bunching curvature radiation and it after the absorption.
We adopt that $\gamma_s=10^2$, $n_{\mathrm{e}}=10^{11}\,\rm cm^{-3}$ and $\rho=10^7$ cm.
The plotted spectra are normalized to the $F_\nu$ of curvature radiation with $\varphi_t=10^{-3}$ at $\omega=\omega_{\mathrm{c}}$.
The spectral index is $5/3$ at lower frequencies due to the self-absorption, which is steeper than pure curvature radiation, but it is still wide-band emission.
Alternatively, the $E_\|$ can let electrons and positrons decouple, which leads to a spectral index of $8/3$ at lower frequencies \citep{2020ApJ...901L..13Y}.
The total results of these independent processes can make a steeper spectrum with an index of $13/3$.

\section{Bunch Expansion}\label{sec4}

\begin{figure}
\resizebox{\hsize}{!}{\includegraphics{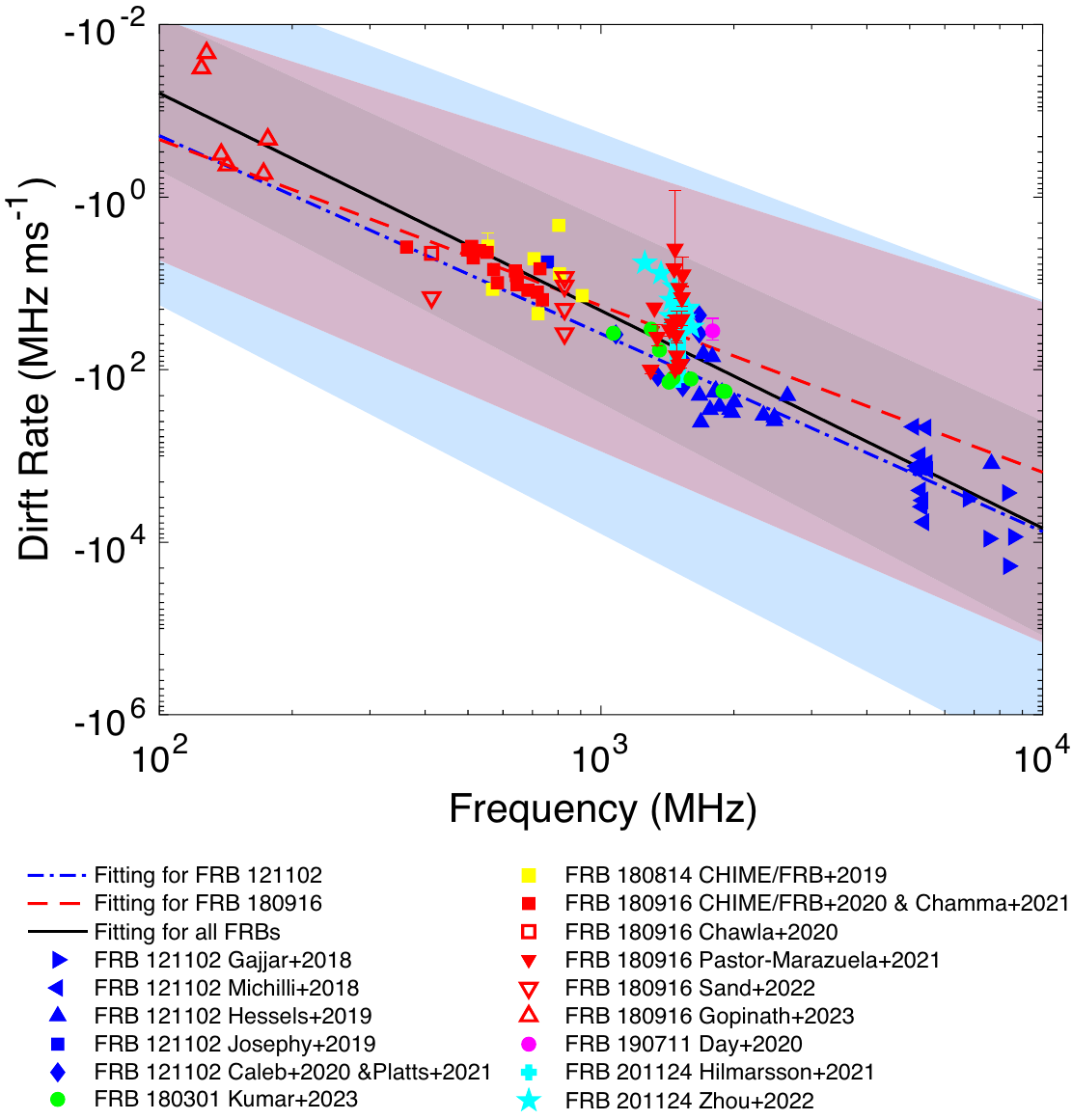}}
\caption{Comparison of drift rates at different frequencies. Different regions show the $1\sigma$ region of the best fitting results: FRB
20121102A (blue), FRB 180916B(red) and all FRBs (grey).
}
\label{fig:nunudot}
\end{figure}

\begin{figure}
\resizebox{\hsize}{!}{\includegraphics{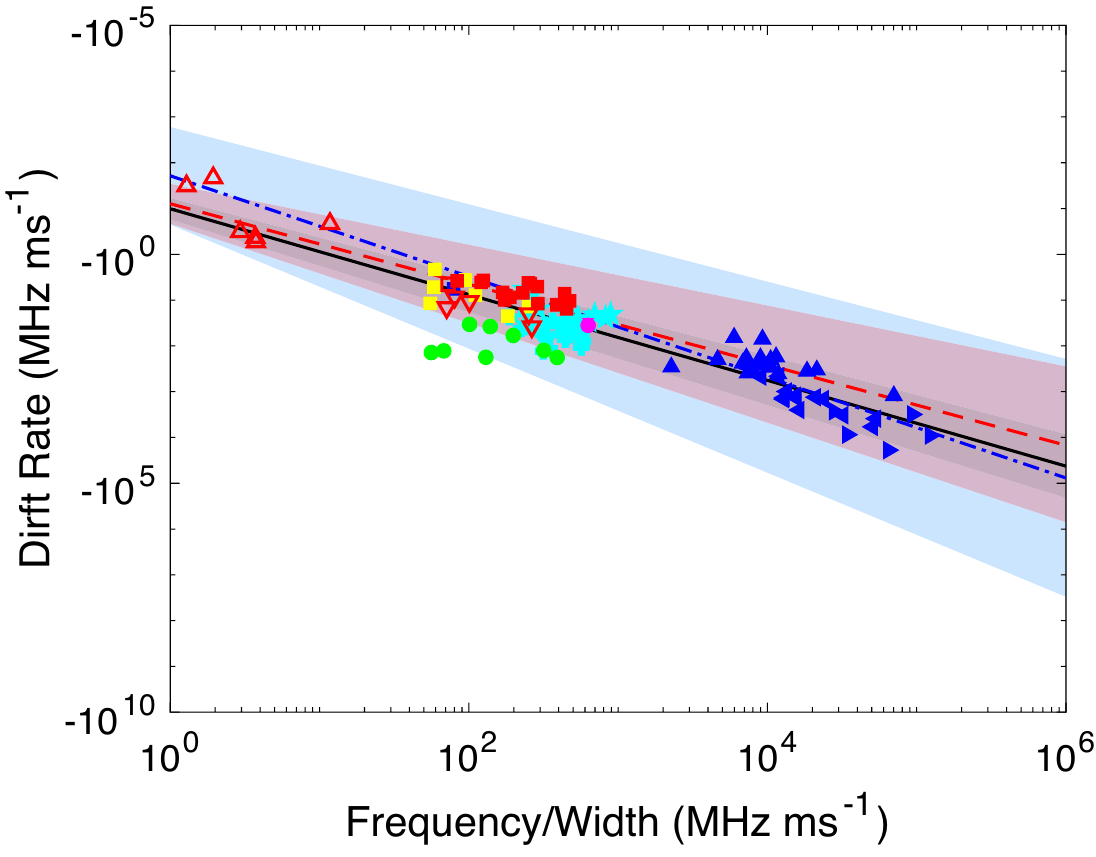}}
\caption{Same as Figure \ref{fig:nunudot} but for comparison of drift rates at different frequency/width.
}
\label{fig:nutnudot}
\end{figure}

As the bunch moves to a higher altitude, the thickness of the bunch (the size parallel to the field lines) becomes larger due to the repulsive force of the same sign charge.
If the bunch travels through the emission region faster than the LOS sweeps across the whole emission beam, the thickness can be estimated as $cw$, where $w$ is the burst width \citep{2022MNRAS.517.5080W}.
For simplicity, we consider the motion of electrons at the boundary due to the Coulomb force from bunches in the co-moving rest frame.
The acceleration of an electron at the boundary is estimated as 
\be
a'_b=\int \frac{e^2n_{\mathrm{e}} l'}{(R_{\rm b}'^2+l'^2)^{3/2}\gamma m_{\mathrm{e}}}dV',
\label{eq:afrombunch}
\ee
where $R_{\rm b}'=R_{\rm b}$ is the transverse size of bunch.
Note that the transverse size of a bunch is much larger than the longitude size even in the co-moving rest frame.
Equation (\ref{eq:afrombunch}) is then calculated as $a'_b\simeq\pi e^2n_{\mathrm{e}}s'/(\gamma m_{\mathrm{e}})$.
Charged particles in the bunch can not move across the $\boldsymbol{B}$-field due to the fast cyclotron cooling.
The transverse size is essentially $R_{\rm b}\propto r$ due to the magnetic freezing, thus, we have $a_b'\propto r^{-2}$.
The distance that the electron travels in the co-moving frame is $s'\propto r$ by assuming $r\gg r_0$ ($r_0$ is an initial distance).
As a result, the thickness of the bunch is $cw\propto r$ in the lab frame.

Burst width $w$ would be larger if the emission is seen from a higher altitude region.
If there is more than one bulk of bunches being observed coincidentally but with different trajectory planes, one would see the drifting pattern \citep{2019ApJ...876L..15W}.
The geometric time delay always lets a burst from a lower region be seen earlier, that is, a downward drifting pattern is observed.
Subpulse structure with such a drifting pattern is a natural consequence caused by the geometric effect when the characteristic frequency of the emission is height-dependent.

Following the previous work \citep{2022ApJ...927..105W}, we continuously investigate the drifting rate as a function of the burst central frequency and width.
The drift pattern is a consequence of narrow-band emission.
Compared with our previous work, we add the fitting results of FRB 20180916B.
Figure \ref{fig:nunudot} and \ref{fig:nutnudot} show the best fitting results for FRB 20121102A (blue dashed-dotted lines), FRB 20180916B (red dashed lines) and all FRBs (black solid lines) and their corresponding $1\sigma$ regions.
The best fitting of $\dot\nu-\nu$ relationship for FRB 20121102A, FRB 20180916B and all FRBs are $(\dot\nu/{\rm MHz\,ms^{-1}})=-10^{-5.30}\times(\nu/{\rm MHz})^{2.29}$,  $(\dot\nu/{\rm MHz\,ms^{-1}})=-10^{-4.52}\times(\nu/{\rm MHz})^{1.93}$, and $(\dot\nu/{\rm MHz\,ms^{-1}})=-10^{-6.25}\times(\nu/{\rm MHz})^{2.52}$.
The best fitting of $\dot\nu$ as a function of $\nu/\rm width$ for FRB 20121102A, FRB 20180916B and all FRBs are $(\dot\nu/{\rm MHz\,ms^{-1}})=-10^{-1.75}\times(\nu/{\rm MHz})^{1.10}(w/{\rm ms})^{-1.10}$, $(\dot\nu/{\rm MHz\,ms^{-1}})=-10^{-1.11}\times(\nu/{\rm MHz})^{0.88}(w/{\rm ms})^{-0.88}$, and $(\dot\nu/{\rm MHz\,ms^{-1}})=-10^{-0.99}\times(\nu/{\rm MHz})^{0.94}(w/{\rm ms})^{-0.94}$.
The fitting results generally match the case of curvature radiation, which has $\dot\nu\propto\nu^2$.
The relationship $\dot\nu\propto\nu w^{-1}$ is attributed to $w\propto r$.
Data are quoted from references as follows: \cite{2018Natur.553..182M,2018ApJ...863....2G,2019ApJ...876L..23H,2019ApJ...882L..18J,2020MNRAS.496.4565C,2021MNRAS.505.3041P,2023MNRAS.526.3652K,2019Natur.566..235C,2020Natur.582..351C,2021MNRAS.507..246C,2020ApJ...896L..41C,2021Natur.596..505P,2022ApJ...932...98S,2020MNRAS.497.3335D,2021MNRAS.508.5354H,2022RAA....22l4001Z,2024MNRAS.527.9872G}.

\section{Discussion}\label{sec6}
\subsection{Curvature Radiation}\label{sec6.1}

Curvature radiation is formed since the perpendicular acceleration on the curved trajectory and its spectrum is naturally wide-band.
Spectra of bunching curvature radiation are more complex due to the geometric conditions and energy distribution (e.g., \citealt{2018ApJ...868...31Y,2022ApJ...927..105W}).
Despite considering the complex electron energy distributions, the spectrum is still hard to be narrow-band.

A quasi-periodically distributed bunches can create such narrow-band emission.
The radiation from the bulk of bunches is quasi-periodically enhanced and the bandwidth depends on $\omega_M$ and the threshold of the telescope.
As shown in Figure \ref{fig:bmspec}, narrow-band emissions with roughly the same bandwidth are expected to be seen from multi-band observations simultaneously.
The energy is radiated into several multiples of $2\pi \omega_M$ with narrow bandwidths and the required number of particles in the bunch is smaller than that of pure curvature radiation.

\subsection{Inverse Compton scattering}\label{sec6.2}

Inverse Compton scattering (ICS) occurs when low-frequency electromagnetic waves enter bunches of relativistic particles \citep{2022ApJ...925...53Z}.
The bunched particles are induced to oscillate at the same frequency in the comoving frame.
The outgoing ICS frequency in the lab-frame is calculated as
\be
\omega\simeq\omega_{\rm in}\gamma^2(1-\cos\theta_{\rm in}),
\label{eq:f-ICS}
\ee
where $\omega_{\rm in}$ is the angular frequency of the incoming wave and $\theta_{\rm in}$ is the angle between the incident photon momentum and the electron momentum.
This ICS process could produce a narrow spectrum for a single electron by giving a low-frequency wave of narrow spectrum \citep{2023Univ....9..375Z}.
In order to generate GHz waves, dozens of kHz incoming wave is required for bunches with $\gamma=10^2$ and $\theta_{\rm in}$ should be larger than $0.5$.
This requires the low-frequency waves to be triggered from a location very far from the emission region.
The X-mode of these waves could propagate freely in the magnetosphere, but O-mode could not unless the waves' propagation path approaches a vacuum-like environment.

As discussed in Section \ref{sec2}, the Lorentz factor is constant due to the balance between $E_\|$ and radiation damping.
$\theta_{\rm in}$ varies slightly as the LOS sweeps across different field lines.
If the incoming wave is monochromatic, the outgoing ICS wave would have a narrow value range matching the observed narrow-band FRB emissions.
The polarization of ICS from a single charge is strongly linearly polarized, however, the electric complex
vectors of the scattered waves in the same direction are
added for all the particles in a bunch.
Therefore, circular polarization would be seen for off-beam geometry \citep{2023MNRAS.522.2448Q}, similar to the case of bunching curvature radiation but they are essentially different mechanisms\citep{2022SCPMA..6589511W}.
Doppler effects may be significant at off-beam geometry so that the burst duration and spectral bandwidth would get larger.
The model predicts that a burst with higher circular polarization components tends to have wider spectral bandwidth.
This may differ from the quasi-periodically distributed bunch case, in which the mechanism to create narrow spectra is independent of polarization.

The perturbation discussed in Section \ref{sec2.2} can let charges oscillate at the same frequency in the comoving frame.
This process is essentially the same as the ICS emission.
The acceleration caused by the perturbation is proposed to be proportional to $\exp(-i\omega_a t)$, which leads to the wave frequency of $\omega_a(1-\beta\cos\theta_{\rm in})$ due to the Doppler effect in the lab frame.
Taking the wave frequency into Equation (\ref{eq:narrowband}), the result is the same as Equation (\ref{eq:f-ICS}).
We let $F(t)=\int E(t)e m^{-1}_{\rm e} {\rm d}t$ and ignore the radiation damping as a simple case.
The dimensionless velocity is then given by $\beta=F(t)/\sqrt{1+F(t)^2}$.
By assuming $F(t)$ has a monochrome oscillating form, according to Equation (\ref{eq:specA5}) if $F(t)\ll1$, the result is reminiscent of the perturbation cases and ICS.
If $F(t)\gg1$, we have $\beta\sim1$ which implies a wide-band emission.

\subsection{Interference Processes}\label{sec6.3}

Some large-scale interference processes, e.g., scintillation, gravitational lensing, and plasma lensing, could change the radiation spectra via wave interference \citep{2023ApJ...956...67Y}. The observed spectra could be coherently enhanced at some frequencies, leading to the modulated narrow spectra. 
We consider that the multipath propagation effect of a given interference process (scintillation, gravitational lensing, and plasma lensing) has a delay time of $\delta t$, then the phase difference between the rays from the multiple images is $\delta\phi\sim2\pi\nu\delta t$, where $\nu$ is the wave frequency. Thus, for GHz waves, a significant spectral modulation requires a delay time of $\delta t\sim10^{-10}\nu_9^{-1}~{\rm s}$, which could give some constraints on scintillation, gravitational lensing, or plasma lensing, if the observed narrow-band is due to these processes.

For scintillation, according to \citet{2023ApJ...956...67Y}, the plasma screen should be at a distance of $\sim10^{15}$ cm from the FRB source, and the medium in the screen should be intermediately dense and turbulent. For plasma lensing, the lensing object needs to have an average electron number density of $\sim10~{\rm cm^{-3}}$ and a typical scale of $\sim10^{-3}$ AU at a distance of $\sim1$ pc from the FRB source. The possibility of gravitational lensing could be ruled out because the gravitational lensing event is most likely a one-time occurrence, meanwhile, it cannot explain the remarkable burst-to-burst variation of the FRB spectra.

\section{Summary}\label{sec7}
In this work, we discuss the spectral properties from the perspectives of radiation mechanisms and absorption effects in the magnetosphere, and the following conclusions are drawn:

1. A narrow spectrum may reflect the periodicity of the distribution of bunches, which have roughly the same Lorentz factor.
The energy is radiated into multiples narrow bands of $2\pi\omega_M$ when the bulk of bunches is quasi-periodic distributed \citep{2023ApJ...956...67Y}.
Such quasi-periodically distributed bunches may be produced due to quasi-monochromatic Langmuir waves or from a ``gap'', which experiences quasi-periodically sparking.
The model predicts that narrow-band emissions may be seen in other higher or lower frequency bands.

2. The quasi-periodically distributed bunches and the perturbation scenarios share the same polarization properties with the uniformly distributed bunches' curvature radiation \citep{2022ApJ...927..105W}.
The emission is generally dominated by linear polarization for the on-beam case whereas it shows highly circular polarization for the off-beam case.
The differential of $V$ is the largest and sign change of circular polarization appears at $\varphi=0$.
The PA across the burst envelope may show slight variations due to the existence of circular polarization.

3. Spectra of bunching curvature radiation with normal distribution are wide-band.
We investigate the spectra from charged bunches with force perturbations on curved trajectories.
If the perturbations are sinusoid with frequency of $\omega_a$, there would be a spike at $2\gamma^2\omega_a$ on the spectra.
However, the perturbations are constrained by the monoenergy of bunches unless the scattering of the Lorentz factor can make the spike wider leading to a wide-band spectrum.
Monochromatic Alfv\'en waves as the perturbations would not give rise to
a narrow-band spectrum from 100 MHz to 10 GHz. 

4. Landau damping as an absorption mechanism may generate a narrow-band spectrum.
We consider three different forms of charge distribution in the magnetosphere.
Subluminous O-mode photons are optically thick, but there is no Landau damping for X-mode and superluminous O-mode photons.
The superluminous O-mode photons have a low-frequency cut-off of $\omega_{\mathrm p}\gamma_{\rm p}^{-3/2}$ which depends on the height in the magnetosphere.
If so, the spectrum would be narrow when $\omega_{\mathrm p}\gamma_{\rm p}^{-3/2}$ is close to $\omega_{\rm c}$ and the bandwidth is getting wider as the frequency becomes lower.
The condition of $\omega_{\rm p}\gamma_{\rm p}^{-3/2}\approx\omega_{\rm c}$ can only be satisfied sometimes.
For some FRBs, the spectrum of some FRBs can be broader than several hundred MHz.
This may be due to the frequency cut-off of the superluminous O-mode being much lower than the characteristic frequency of curvature radiation, or the waves propagating via X-mode.

5. Self-absorption by curvature radiation is significant at lower frequencies.
By considering the Einstein coefficients, the cross section is larger for lower frequencies.
The spectral index is $5/3$ steeper than that of pure curvature radiation at $\omega\ll\omega_{\rm c}$, however, the spectrum is not as narrow as $\Delta\nu/\nu=0.1$.

6. The two intrinsic mechanisms by invoking quasi-periodic bunch distribution and the perturbations are more likely to generate the observed narrow spectra of FRBs rather than the absorption effects.

7. As a bulk of bunches moves to a higher region, the longitude size of the bulk becomes larger, suggesting that burst duration is larger for bulks emitting at higher altitudes.
By investigating drifting rates as functions of the central frequency and $\nu/w$, we find that drifting rates can be characterized by power law forms in terms of the central frequency and $\nu/w$.
The fitting results are generally consistent with curvature radiation.

\begin{acknowledgements}
We are grateful to the referee for constructive comment and V. S. Beskin, He Gao, Biping Gong, Yongfeng Huang, Kejia Lee, Ze-Nan Liu, Jiguang Lu, Rui Luo, Chen-Hui Niu, Jiarui Niu, Yuanhong Qu, Hao Tong, Shuangqiang Wang, CHengjun Xia, Zhenhui Zhang, Xiaoping Zheng, Enping Zhou, Dejiang Zhou, and Yuanchuan Zou for helpful discussion.
W.Y.W. is grateful to Bing Zhang for the discussion of the ICS mechanism and his encouragement when I feel confused about the future.
This work is supported by the National SKA Program of China (No. 2020SKA0120100) and the Strategic Priority Research Program of the CAS (No. XDB0550300).
Y.P.Y. is supported by the National Natural Science Foundation of China (No.12003028) and the National SKA Program of China (2022SKA0130100).
J.F.L. acknowledges support from the NSFC (Nos.11988101 and 11933004) and from the New Cornerstone Science Foundation through the New Cornerstone Investigator Program and the XPLORER PRIZE.
\end{acknowledgements}

\appendix\label{appendix}
\section{Bunch Radiation}\label{app:radiation}
In order to obtain the spectrum from bunches, we briefly summarize the physics of radiation from a single charge.
As shown in Figure \ref{fig:bunch}, the angle between the electron velocity direction and $x$-axis at $t = 0$ is defined as $\chi_{ij}$, and that between the LOS and the trajectory plane is $\varphi_k$.
The observation point is far enough away from
the region of space where the acceleration occurs.
Based on the classical electrodynamics, the radiation electric vector is given by \citep{1998clel.book.....J}
\be
\boldsymbol{E}(t)=\frac{e}{c}\left\{\frac{\boldsymbol{n}\times[(\boldsymbol{n}-\boldsymbol{\beta}_s)\times\boldsymbol{\dot\beta}_s]}{(1-\boldsymbol{\beta}_s\cdot\boldsymbol{n})^3\mathcal{R}}\right\},
\label{eq:radiationEvector}
\ee
where the expression in the brackets is evaluated at the retarded time $t_{\rm ret}=t-\mathcal{R}/c$.
By inserting the two accelerations $\dot{\beta}_\|$ and $\dot{\beta}_\perp$, the vector component of Equation (\ref{eq:radiationEvector}) has (the subscript `$i,j,k$' has been replaced by `$s$')
\be
\begin{array}{c}
\boldsymbol{n}\times[(\boldsymbol{n}-\boldsymbol{\beta}_s)\times\boldsymbol{\dot\beta}_s]\\
=-\dot{\beta}_{s\|} \sin (\chi_s+v_st/\rho)\boldsymbol{\epsilon}_1+\dot{\beta}_{s\perp}[\cos \varphi_s-\cos (\chi_s+v_st/\rho)]\boldsymbol{\epsilon}_1\\
+\dot{\beta}_{s\|} \sin \varphi_s\cos (\chi_s+v_st/\rho)\epsilon_2-\dot{\beta}_{s\perp} \sin \varphi_s \sin (\chi_s+v_st/\rho) \boldsymbol{\epsilon}_2.
\label{eq:betadot}
\end{array}
\ee
Consider the small value of $\varphi_s$, $\chi_s$ and $v_s t/\rho$, if $\dot{\beta}_{s\|}\simeq\dot{\beta}_{s\perp}$, the radiation power caused by $\dot{\beta}_{s\perp}$ is much smaller than that for $\dot{\beta}_{s\|}$.

The energy per unit frequency per unit solid angle of a single particle is given by
\be
\begin{aligned}
\frac{{\rm d}^2W}{{\rm d}\omega {\rm d}\Omega}&=\frac{c}{4\pi^2}\left|\mathcal{R}\boldsymbol{E}_s(t)\exp(i\omega t){\rm d}t\right|^2\\
&=\frac{e^2}{4 \pi^2 c}\left|\int_{-\infty}^{\infty} \frac{\boldsymbol{n} \times[(\boldsymbol{n}-\boldsymbol{\beta}_s) \times \dot{\beta}_s]}{(1-\boldsymbol{\beta}_s \cdot \boldsymbol{n})^2}\right.\\
&\left.\times\exp[i \omega(t'-\boldsymbol{n} \cdot \boldsymbol{r}_s(t') / c)]{\rm d}t'\right|^2.
\end{aligned}
\label{eq:specA3}
\ee
Using the identity:
\begin{equation}
\frac{\boldsymbol{n} \times[(\boldsymbol{n}-\boldsymbol{\beta}_s) \times \dot{\boldsymbol{\beta}}_s]}{(1-\boldsymbol{\beta}_s \cdot \boldsymbol{n})^2}=\frac{{\rm d}}{{\rm d} t'}\left[\frac{\boldsymbol{n} \times(\boldsymbol{n} \times \boldsymbol{\beta}_s)}{1-\boldsymbol{\beta}_s \cdot \boldsymbol{n}}\right],
\end{equation}
Equation (\ref{eq:specA3}) becomes
\be
\frac{{\rm d}^2W}{{\rm d}\omega {\rm d}\Omega} =\frac{e^2}{4 \pi^2 c}\left|\int_{-\infty}^{\infty} \exp[i \omega(t'-\boldsymbol{n} \cdot \boldsymbol{r}_s(t') / c)]{\rm d}\left[\frac{\boldsymbol{n} \times(\boldsymbol{n} \times \boldsymbol{\beta}_s)}{1-\boldsymbol{\beta}_s \cdot \boldsymbol{n}}\right]\right|^2.
\label{eq:specA4}
\ee
It can be proven that by adding or subtracting the integrals over the times while maintaining a constant velocity from Equation (\ref{eq:specA4}), one can use partial integration to obtain
\be
\frac{{\rm d}^2W}{{\rm d}\omega {\rm d}\Omega} =\frac{e^2\omega^2}{4 \pi^2 c}\left|\int_{-\infty}^{\infty} \boldsymbol{n} \times(\boldsymbol{n} \times \boldsymbol{\beta}_s)\exp[i \omega(t'-\boldsymbol{n} \cdot \boldsymbol{r}_s(t') / c)]{\rm d}t'\right|^2,
\label{eq:specA5}
\ee
as long as a convergence factor (e.g., $\exp(-\epsilon|t|)$, where taking the limit $\epsilon\rightarrow0$) is inserted to the integrand function of Equation (\ref{eq:specA4}) to eliminate instability at $t=\pm\infty$.

The energy radiated per unit solid angle per unit frequency interval in terms of the two polarization components can be written as
\begin{equation}
\frac{{\rm d}^2W}{{\rm d}\omega {\rm d}\Omega} =\frac{e^2 \omega^2}{4 \pi^2 c}\left|-\boldsymbol{\epsilon}_{\|} A_{\|}+\boldsymbol{\epsilon}_{\perp} A_{\perp}\right|^2.
\label{eq:AparaAperp}
\end{equation}
As shown in Figure \ref{fig:bunch}, we consider a charge of identifier of $ijk$.
The vector part of Equation (\ref{eq:specA5}) can be written as
\be
\begin{aligned} &\Vec{n}\times\left(\boldsymbol{n} \times \boldsymbol{\beta}_{ijk,\|}\right)\\
&=\beta_\|\left[-\boldsymbol{\epsilon}_{\|} \sin \left(\frac{v t}{\rho}+\chi_{ij}\right)+\boldsymbol{\epsilon}_{\perp} \cos \left(\frac{v t}{\rho}+\chi_{ij}\right) \sin \varphi_k\right],
\end{aligned}
\ee
and the argument of the exponential is
\begin{equation}
\begin{aligned}
\omega\left(t-\frac{\boldsymbol{n} \cdot \boldsymbol{r}_{ijk}(t)}{c}\right)
&=\omega\left[t-\frac{2 \rho}{c} \sin \left(\frac{v t}{2 \rho}\right) \cos \left(\frac{v t}{2 \rho}+\chi_{ij}\right) \cos \varphi_k\right]\\
&\simeq \frac{\omega}{2}\left[\left(\frac{1}{\gamma^{2}}+\varphi_k^{2}+\chi_{ij}^{2}\right) t+\frac{c^{2} t^{3}}{3 \rho^{2}}+\frac{c t^{2}}{\rho} \chi_{ij}\right],
\end{aligned}
\end{equation}
where $t$ is adopted to replace $t'$ in the following calculations.
Therefore, the two amplitudes for Equation (\ref{eq:AparaAperp}) are given by
\be
\begin{aligned}
A_{ijk,\|} &\simeq \int_{-\infty}^{\infty}\left(\frac{c t}{\rho}+\chi_{ij}\right)\\
&\times\exp \left(i \frac{\omega}{2}\left[\left(\frac{1}{\gamma^{2}}+\varphi_k^{2}+\chi_{ij}^{2}\right) t+\frac{c^{2} t^{3}}{3 \rho^{2}}+\frac{c t^{2}}{\rho} \chi_{ij}\right]\right){\rm d}t, \\
A_{ijk,\perp} &\simeq \varphi_k \int_{-\infty}^{\infty} \exp \left(i \frac{\omega}{2}\left[\left(\frac{1}{\gamma^{2}}+\varphi_k^{2}+\chi_{ij}^{2}\right) t+\frac{c^{2} t^{3}}{3 \rho^{2}}+\frac{c t^{2}}{\rho} \chi_{ij}\right]\right){\rm d}t .
\end{aligned}
\label{eq:specA6}
\ee
Making the changes of variables
\begin{equation}
\begin{aligned}
&u=\frac{c t}{\rho}\left(\frac{1}{\gamma^{2}}+\varphi_{k}^{2}+\chi_{ij}^2\right)^{-1 / 2}, \\
&\xi=\frac{\omega \rho}{3 c}\left(\frac{1}{\gamma^{2}}+\varphi_{k}^{2}+\chi_{ij}^2\right)^{3 / 2},
\end{aligned}
\label{eq:uxi}
\end{equation}
Equation (\ref{eq:specA6}) becomes
\begin{equation}
\begin{aligned}
A_{ijk,\|}& \simeq \frac{\rho}{c}\left(\frac{1}{\gamma^{2}}+\varphi_{k}^2+\chi_{ij}^2\right)\\
&\times\int_{-\infty}^{\infty}\left(u+\frac{\chi_{ij}}{\sqrt{1 / \gamma^{2}+\chi_{ij}^{2}+\varphi_k^2}}\right) \\
&\times\exp \left[i \frac{3}{2} \xi\left(u+\frac{1}{3} u^{3}+\frac{\chi_{ij}}{\sqrt{1 / \gamma^{2}+\varphi_k^2+\chi_{ij}^{2}}} u^{2}\right)\right] {\rm d} u,\\
A_{ijk,\perp}&\simeq
\frac{\rho}{c}\varphi_k\left(\frac{1}{\gamma^{2}}+\varphi_{k}^2+\chi_{ij}^2\right)^{1/2} \\
&\times\int_{-\infty}^{\infty}\exp \left[i \frac{3}{2} \xi\left(u+\frac{1}{3} u^{3}+\frac{\chi_{ij}}{\sqrt{1 / \gamma^{2}+\varphi_k^2+\chi_{ij}^{2}}} u^{2}\right)\right] {\rm d} u.\\
\end{aligned}
\label{eq:specA7}
\end{equation}
Note that
\be
\begin{aligned}
&\lim\limits_{u\to+0} \left(u+\frac{1}{3} u^{3}+\frac{\chi_{ij}}{\sqrt{1 / \gamma^{2}+\varphi_k^2+\chi_{ij}^{2}}} u^{2}\right)=u,\\
&\lim\limits_{u\to+\infty} \left(u+\frac{1}{3} u^{3}+\frac{\chi_{ij}}{\sqrt{1 / \gamma^{2}+\varphi_k^2+\chi_{ij}^{2}}} u^{2}\right)=\frac{1}{3}u^3.
\end{aligned}
\ee
The integrals in Equation (\ref{eq:specA7}) are identifiable as Airy integrals. Consequently, the two amplitudes are
\begin{equation}
\begin{aligned}
A_{ijk,\|} &\simeq \frac{i2}{\sqrt{3}}\frac{\rho}{c}\left(\frac{1}{\gamma^{2}}+\varphi_{k}^2+\chi_{ij}^2\right)K_{\frac{2}{3}}(\xi)\\
&+ \frac{2}{\sqrt{3}}\frac{\rho}{c}\chi_{ij}\left(\frac{1}{\gamma^{2}}+\varphi_{k}^2+\chi_{ij}^2\right)^{1/2}K_{\frac{1}{3}}(\xi),\\
A_{ijk,\perp} &\simeq \frac{2}{\sqrt{3}}\frac{\rho}{c}\varphi_k\left(\frac{1}{\gamma^{2}}+\varphi_{k}^2+\chi_{ij}^2\right)^{1/2}K_{\frac{1}{3}}(\xi),
\end{aligned}
\label{eq:specA8}
\end{equation}
where $K_\nu(\xi)$ is the modified Bessel function, in which $K_\nu(\xi) \rightarrow(\Gamma(\nu) / 2)(\xi / 2)^{-\nu}$ for $\xi \ll 1$ and $\nu \neq 0$, and $K_\nu(\xi) \rightarrow \sqrt{\pi /( 2 \xi)} \exp (-\xi)$ for $\xi\gg1$, leading to a wide-band spectrum.

The curvature radius can be regarded as a constant since the bunch's size is much smaller than its altitude.
We assume that each bunch has roughly the same $\chi_u$ and $\chi_d$ even with different $k$ so that an average height can be derived to replace the complex boundary condition of the bulk \citep{2022ApJ...927..105W}.
Consider that the bunch longitude size is $l<4\rho/(3\gamma^3)$.
The emission with $\omega<\omega_{\mathrm{c}}$ is added coherently.
For the emission band with $\omega>\omega_l$, the emission is added incoherently and the fluxes drop with $\omega$ very quickly.
The spectrum becomes much deeper than an exponential function.
The total amplitudes of a bunch are then given by
\be
\begin{aligned}
A_{\|}&\simeq\frac{2}{\sqrt{3}}\frac{\rho}{c}\frac{N_j}{\Delta\chi}\frac{N_k}{2\varphi_t}N_i\\
&\times\int^{\chi_{u,i}}_{\chi_{d,i}}d\chi'\int^{\varphi_u}_{\varphi_d}\left[i\left(\frac{1}{\gamma^{2}}+\varphi'^2+\chi'^2\right)K_{\frac{2}{3}}(\xi)\right.\\
&\left.+ \chi'\left(\frac{1}{\gamma^{2}}+\varphi'^2+\chi'^2\right)^{1/2}K_{\frac{1}{3}}(\xi)\right]\cos\varphi'{\rm d}\varphi',\\
A_{\perp}& \simeq \frac{2}{\sqrt{3}}\frac{\rho}{c}\frac{N_j}{\Delta\chi}\frac{N_k}{2\varphi_t}N_i\\
&\times\int^{\chi_{u,i}}_{\chi_{d,i}}d\chi'\int^{\varphi_u}_{\varphi_d}\left(\frac{1}{\gamma^{2}}+\varphi'^2+\chi'^2\right)^{1/2}K_{\frac{1}{3}}(\xi)\varphi'\cos\varphi'{\rm d}\varphi',
\end{aligned}
\label{eq:bunchA}
\ee
where $\chi_u$ is the upper boundary $\chi'$, $\chi_d$ is the lower boundary $\chi'$, and the boundaries of $\varphi'$ read $\varphi_u=\varphi_t+\varphi$ and $\varphi_d=-\varphi_t+\varphi$.

With the particular interest of $\beta_{1,\|}(t)=\beta_1\exp(-i\omega_at)$, the total amplitudes of a bunch for the parallel perturbation are given by
\be
\begin{aligned}
\tilde{A}_{\|}&\simeq\frac{2}{\sqrt{3}}\frac{\rho}{c}\frac{N_j}{\Delta\chi}\frac{N_k}{2\varphi_t}N_i\beta_1\\
&\times\int^{\chi_{u,i}}_{\chi_{d,i}}d\chi'\int^{\varphi_u}_{\varphi_d}\left[i\left(\frac{1}{\gamma^{2}}+\varphi'^2+\chi'^2-\frac{2\omega_a}{\omega}\right)K_{\frac{2}{3}}(\xi_a)\right.\\
&\left.+\chi'\left(\frac{1}{\gamma^{2}}+\varphi'^2+\chi'^2-\frac{2\omega_a}{\omega}\right)^{1/2}K_{\frac{1}{3}}(\xi_a)\right]\cos\varphi'{\rm d}\varphi',\\
\tilde{A}_{\perp} &\simeq \frac{2}{\sqrt{3}}\frac{\rho}{c}\frac{N_j}{\Delta\chi}\frac{N_k}{2\varphi_t}N_i\beta_1\\
&\times\int^{\chi_{u,i}}_{\chi_{d,i}}d\chi'\int^{\varphi_u}_{\varphi_d}\left(\frac{1}{\gamma^{2}}+\varphi'^2+\chi'^2-\frac{2\omega_a}{\omega}\right)^{1/2}\\
&\times K_{\frac{1}{3}}(\xi_a)\varphi'\cos\varphi'{\rm d}\varphi'.
\end{aligned}
\label{eq:bunchA1}
\ee
The monochromatic perturbation frequency is Doppler boosted to $2\gamma^2\omega_a$ so that there is a spike in the spectrum at $\omega=2\gamma^2\omega_a$, and the width of the spike is determined by the boundaries of $\chi'$ and $\varphi'$.

We consider the case with a similar formula but for the perpendicular perturbation, i.e., $\beta_{1,\perp}(t)=\beta_1\exp(-i\omega_at)$.
Following the discussion in Section \ref{sec2.2.4}, we insert Equation (\ref{eq:newuxi}) into Equation (\ref{eq:AmplitudePer}) and the amplitudes are calculated as
\be
\begin{aligned}
\tilde{A}_{ijk,\|}&\simeq\frac{1}{\sqrt{3}}\frac{\rho}{c}\beta_1(\varphi_k^2-\chi_{ij}^2)\left(\frac{1}{\gamma^{2}}+\varphi'^2+\chi'^2-\frac{2\omega_a}{\omega}\right)^{1/2}K_{\frac{1}{3}}(\xi_a)\\
&-\frac{i2}{\sqrt{3}}\frac{\rho}{c}\beta_1\chi_{ij}\left(\frac{1}{\gamma^{2}}+\varphi'^2+\chi'^2-\frac{2\omega_a}{\omega}\right)K_{\frac{2}{3}}(\xi_a)\\
&+\frac{\rho}{c}\beta_1\left(\frac{1}{\gamma^{2}}+\varphi'^2+\chi'^2-\frac{2\omega_a}{\omega}\right)^{3/2}\\
&\times\left[\frac{1}{\sqrt{3}}K_{\frac{1}{3}}(\xi_a)-\pi\delta\left(3\xi_a/2\right)\right],\\
\tilde{A}_{ijk,\perp}&\simeq -\frac{i2}{\sqrt{3}}\frac{\rho}{c}\varphi_k\beta_1\left(\frac{1}{\gamma^{2}}+\varphi_{k}^2+\chi_{ij}^2\right)K_{\frac{2}{3}}(\xi_a)\\
&- \frac{2}{\sqrt{3}}\frac{\rho}{c}\varphi_k\beta_1\chi_{ij}\left(\frac{1}{\gamma^{2}}+\varphi_{k}^2+\chi_{ij}^2\right)^{1/2}K_{\frac{1}{3}}(\xi_a),\\
\end{aligned}
\label{eq:bunchA2}
\ee
where $\delta(x)$ is the Dirac function. 
The total amplitudes of a bunch for the perpendicular perturbation case can be given by the integrals of Equation (\ref{eq:bunchA2}).

\section{Stokes Parameters}\label{app:stokes}
Four Stokes parameters are significant tools to describe the polarization properties of a quasi-monochromatic electromagnetic wave.
In this section, we consider the monochrome Stokes parameters which can be derived by the Fourier-transformed electric vector.
The Fourier-transformed polarized components of electric waves can be written as
\be
E_{\|}(\omega)=\frac{e\omega A_{\|}}{2\pi c\mathcal{R}},\,
E_{\perp}(\omega)=\frac{e\omega A_{\perp}}{2\pi c\mathcal{R}}.
\label{eq:AtoE}
\ee
By inserting Equation (\ref{eq:AtoE}), four Stokes parameters of observed waves can be calculated as
\begin{equation}
\begin{aligned}
&I=\mu\left(A_{\|} A_{\|}^{*}+A_{\perp} A_{\perp}^{*}\right),\\
&Q=\mu\left(A_{\|} A_{\|}^{*}-A_{\perp} A_{\perp}^{*}\right), \\
&U=\mu\left(A_{\|} A_{\perp}^{*}+A_{\perp}A_{\|}^{*}\right),\\
&V=-i\mu\left(A_{\|}A_{\perp}^{*}-A_{\perp}A_{\|}^{*}\right),
\end{aligned}
\label{eq:stokes}
\end{equation}
where $\mu=\omega^2 e^2/(4\pi^2 \mathcal{R}^2 c \mathcal{T})$ is the proportionality factor, $I$ defines the total intensity, $Q$ and $U$ define linear polarization and its position angle, and $V$ describes circular polarization.
The timescale {$\mathcal{T}$} is chosen as the average timescale the flux density $I$ repeated.
The corresponding linear polarization and PA are then given by
\be
\begin{aligned}
& L=\sqrt{U^2+Q^2},\\
&\Psi=\frac{1}{2} \tan ^{-1}\left(\frac{U_s}{Q_s}\right),
\label{eq:PA}
\end{aligned}
\ee
where
\begin{equation}
\left(\begin{array}{l}
U_s \\
Q_s
\end{array}\right)=\left(\begin{array}{cc}
\cos 2\psi & \sin 2\psi \\
-\sin 2\psi & \cos 2\psi
\end{array}\right)\left(\begin{array}{l}
U \\
Q
\label{eq:pa}
\end{array}\right),
\end{equation}
in which $\psi$ is given by the rotation vector model (RVM; \citealt{1969ApL.....3..225R}).

\section{Quake-induced Oscillation}\label{app:quake}

We calculate the frequencies of the torsional modes for the relativistic star caused by a quake and discuss the effect of the magnetic field on the frequencies of the various torsional modes.
Oscillations after the quake have two types: spherical and toroidal, in which only the toroidal oscillations can effectively modify the spin of the star and the Goldreich-Julian density.
The axial perturbation equations for
the elastic solid star in the Cowling approximation is written as
\citep{Samuelsson:2006tt, Sotani:2012qc}
\begin{equation}
\begin{aligned}
&Y\left[ \frac{\rho_m+P}{\mu}\,\Omega^2 \exp(-2\Phi)-\frac{(\ell+2)(\ell-1)}{r^2}\right]\exp({2\Lambda}) +Y''\\
&  +  \left( \frac{4}{r}+\Phi'-\Lambda'+\frac{\mu'}{\mu} \right) Y' = 0 ,  
\end{aligned}
\label{eq: fc_1}
\end{equation}
where $\mu$ is the shear modulus, $\Phi$ and $\Lambda$ are functions of $r$, $\rho_m$ is the mass-energy density, $Y(r)$ describes the radial part of the angular oscillation amplitude, and the integer $\ell$ is the angular separation constant which enters when $Y(r)$ is expanded in spherical harmonics $Y_{\ell m} (\theta, \phi)$.

To discuss the possible effects of the magnetic field, 
we extend our studies to the torsional oscillation of a magnetized
relativistic star (e.g., \citealt{2024MNRAS.527..855L}). \citet{Sotani:2006at} derived the perturbation equations of
the magnetized relativistic star using the relativistic Cowling approximation.
The final perturbation equation is
\begin{equation}
A_{\ell}(r)Y'' +B_{\ell}(r)Y'+C_{\ell}(r)Y=0 \,,
\label{eq: fc_2}
\end{equation}
where the coefficients are given in terms of the functions describing the
equilibrium metric, fluid, and the magnetic field of the star,
\begin{align}
 A_{\ell}(r) &= \mu + (1 + 2 \lambda_1)\frac{{a_1}^2}{\pi r^4}, 
 \label{eq: coefficients_1} \\
B_{\ell}(r) &=\left(\frac{4}{r} + \Phi' - \Lambda'\right)\mu + \mu'  \nonumber\\ 
      & + (1 + 2\lambda_1)\frac{a_1}{\pi r^4}\left[\left(\Phi' - \Lambda'\right)a_1
     + 2{a_1}'\right] ,
\label{eq: coefficients_2} \\
C_{\ell}(r) &= (2 + 5\lambda_1)\frac{a_1}{2\pi r^4}\left\{\left(\Phi'
- \Lambda'\right){a_1}' + {a_1}''\right\} \nonumber\\
&- (\lambda-2)\left(\frac{ \mu \exp(2\Lambda)}{r^2}
- \frac{\lambda_1{{a_1}'}^2}{2\pi r^4}\right) \nonumber\\
& +\left[\left(\rho_m + P + (1 +2\lambda_1)\frac{{a_1}^2}{\pi r^4}
\right)\exp(2\Lambda)
     - \frac{\lambda_1 {{a_1}'}^2}{2\pi r^2}\right]\\
  &  \times \Omega^2 \exp(-2\Phi)  \nonumber,
\label{eq: coefficients_3} 
\end{align}
where $ \lambda = \ell(\ell+1)$, and $\lambda_1 = -
\ell(\ell+1)/[(2\ell-1)(2\ell+3)]$.  To solve Eqs.~(\ref{eq: fc_1}) and
(\ref{eq: fc_2}) and determine the oscillation frequencies, the boundary
conditions require that the traction vanishes at the top and the bottom of the
crust. 

The toroidal oscillations can effectively modify the spin and the electric field in the gap.
The velocity components of toroidal oscillations can be written
as
\begin{equation}
\delta v^{\hat{i}}=\left[0, \frac{1}{\sin \theta} \partial_\phi Y_{\ell m}(\theta, \phi),-\partial_\theta Y_{\ell m}(\theta, \phi)\right] \tilde{\eta}(r) e^{-i \Omega t},
\end{equation}
where $\tilde{\eta}(r)$ is a parameter denoting the amplitude of the oscillations.
Owing to unipolar induction, we have \citep{2015ApJ...799..152L}
\begin{equation}
E_\theta= -\frac{1}{N c}\left[\frac{2\pi r}{P} \sin \theta\left(1-\frac{R^3\kappa}{r^3}\right)-\partial_\theta Y_{\ell m} \tilde{\eta}(r)\right]
 B_{\rm s} \frac{f(r)}{f(R)} \frac{R^3\cos \theta}{r^3},
\end{equation}
where $N=\sqrt{1-2GM/(rc^2)}$, $\kappa=2GI_m/(R^3c^2)$, $I_m$ is the moment of inertia, and $f(r)$ is
\begin{equation}
f(r)=-3\left(\frac{rc^2}{2GM}\right)^3\left[\ln \left(1-\frac{2GM}{rc^2}\right)+\frac{2GM}{rc^2}\left(1+\frac{GM}{ rc^2}\right)\right].
\end{equation}

\section{Dispersion Relationship of Ultra-relativistic Plasma}\label{app:dispersion}

\begin{figure}
\resizebox{\hsize}{!}{\includegraphics{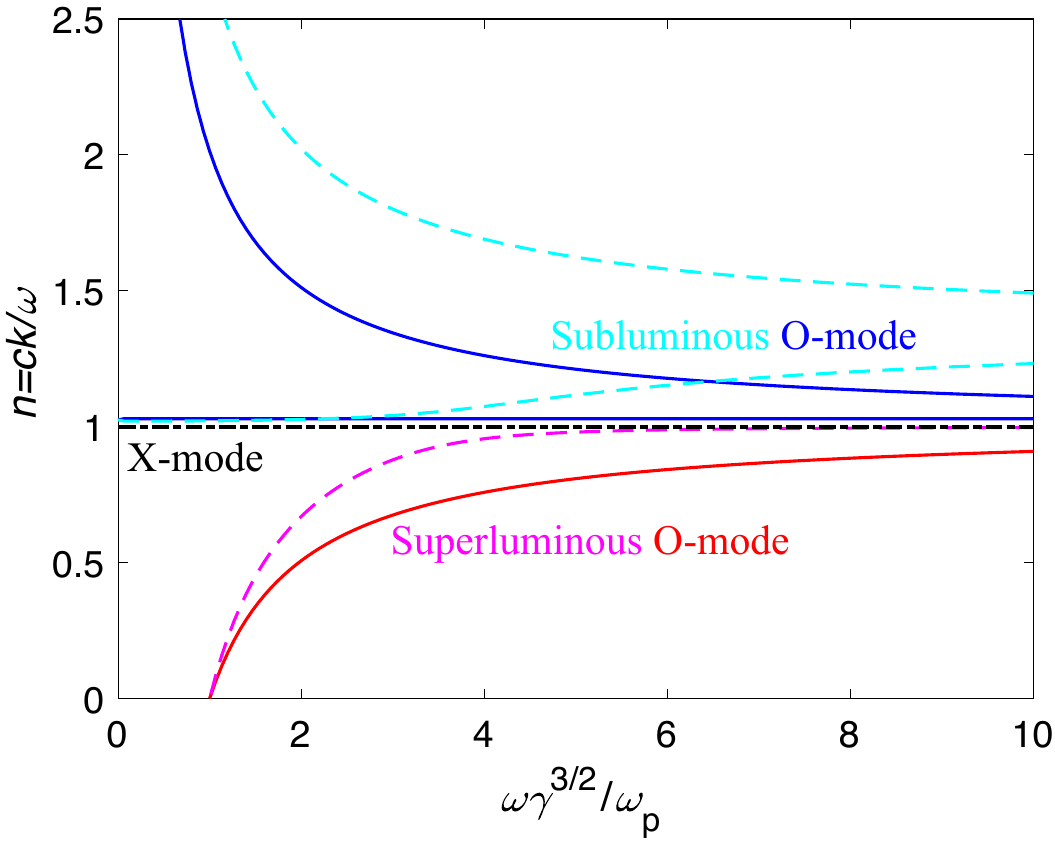}}
\caption{The refractive indices of wave modes as a function of $\omega\gamma^3/\omega_{\mathrm{p}}$ with $\theta_B=0.2$ for two cases: $\gamma=2$ (dashed lines) and $\gamma=100$ (solid lines). The black dotted-dashed line shows the X-mode ($n=1$). The O-modes have two branches: subluminous O-mode (cyan dashed line for $\gamma=2$ and blue solid line for $\gamma=10$) and superluminous O-mode (magenta dashed line for $\gamma=2$ and red solid line for $\gamma=10$).
}
\label{fig:modes}
\end{figure}

In this section, we briefly summarize the physics of waves in ultra-relativistic plasma.
Basically, the equation of motion of a given charge species reads
\begin{equation}
\frac{{\rm d}}{{\rm d} t}\left(\gamma_s m_{\mathrm{e}} \boldsymbol{v}_s\right)=q_s \boldsymbol{E}+\frac{q_s}{c} \boldsymbol{v}_s \times \boldsymbol{B},
\label{eq:dynamic}
\end{equation}
where $\boldsymbol{v}_s=\boldsymbol{v}_{s, 0}+\delta \boldsymbol{v}_s, \boldsymbol{E}=\delta \boldsymbol{E}, \boldsymbol{B}=\boldsymbol{B}_0+\delta \boldsymbol{B}$, in which
$\delta \boldsymbol{v}_s, \delta \boldsymbol{E}, \delta \boldsymbol{B}$ are associated with the disturbance in the plasma and have a form of $\exp{[-i(\omega t-\boldsymbol{k}\cdot \boldsymbol{r})]}$.
Combined with 
\begin{equation}
{\rm d} \delta \boldsymbol{v}_s / {\rm d} t=-i \omega \delta \boldsymbol{v}_s+i c\left(\boldsymbol{k} \cdot \boldsymbol{v}_{s, 0}\right) \delta \boldsymbol{v}_s,
\end{equation}
and
\begin{equation}
\delta \boldsymbol{B}=(c \boldsymbol{k} / \omega) \times \delta \boldsymbol{E},
\end{equation}
$\delta \boldsymbol{v}_s$ can be then described in terms of $\delta \boldsymbol{E}$.
The continuity equation reads
\be
\frac{\partial n_s}{\partial t}+\nabla\cdot(n_s\boldsymbol{v}_{s,0})=0, 
\ee
so that the density perturbation is given by
\be
\delta n_s=\frac{n_s(\boldsymbol{k}\cdot\delta \boldsymbol{v}_s)}{\omega-\boldsymbol{k}\cdot\boldsymbol{v}_{s,0}}.
\ee
The conductivity tensor $\boldsymbol{\sigma}$ is defined by
\be
\delta \boldsymbol{J}=\boldsymbol{\sigma}\cdot\delta\boldsymbol{E}=\Sigma_s(n_sq_s\delta \boldsymbol{v}_s+\delta n_sq_s\boldsymbol{v}_{s,0}),
\label{eq:conductivitytensor}
\ee
and the dielectric tensor is given by
\be
\boldsymbol{\epsilon}=\boldsymbol{I}+i\frac{4\pi}{\omega}\boldsymbol{\sigma},
\label{eq:dielectrictensor}
\ee
where $\boldsymbol{I}$ is the unit tensor. Since the electric displacement vector is defined as $\boldsymbol{D}=\boldsymbol{\epsilon}\cdot\boldsymbol{E}$, we can obtain the equation for plane waves in a coordinate system ($x',\, y',\, z'$) where $\boldsymbol{B_0}$ along $z'$
\be
\epsilon=
\left(\begin{array}{ccc}
S & iD & A \\
-iD & S & -iC \\
A & iC & P
\end{array}\right),
\ee
with
\be
\begin{aligned}
S & =1+\sum_s f_{s, 11}, \\
D & =\sum_s f_{s, 12}, \\
A & =\sum_s \zeta_s f_{s, 11}, \\
C & =\sum_s \zeta_s f_{s, 12}, \\
P & =1+\sum_s\left(f_{s, \eta}+\zeta_s^2 f_{s, 11}\right),
\end{aligned}
\ee
where
\be
\begin{aligned}
f_{s, 11} & =-\frac{\omega_{\mathrm{p}}^2 \gamma_s^{-1}}{\omega^{2}-\omega_B^2 \gamma_s^{-2}\left(1-n \beta_s \cos \theta_B\right)^{-2}}, \\
f_{s, 12} & =-\frac{\operatorname{sign}\left(q_s\right) \omega_B\omega_{\mathrm{p}}^2\omega^{-3} \gamma_s^{-2}\left(1-n \beta_s \cos \theta_B\right)^{-1}}{1-\omega_B^2\omega^{-2} \gamma_s^{-2}\left(1-n \beta_s \cos \theta_B\right)^{-2}}, \\
f_{s, \eta} & =-\frac{\omega_{\mathrm{p}}^2}{\omega^{2}\gamma_s^3\left(1-n \beta_s \cos \theta_B\right)^2}, \\
\zeta_s & =\frac{n \beta_s \sin \theta_B}{1-n \beta_s \cos \theta_B},
\end{aligned}
\ee
here the subscript `0' has been suppressed. In the coordinate system $x_0y_0z_0$ with $k$ along the $z_0$-axis and $B$ in the $x_0-z_0$ plane, the components of dielectric tensor are given by
\be
\begin{aligned}
\epsilon_{x x} & =S \cos ^2 \theta_B-2 A \sin \theta_B \cos \theta_B+P \sin ^2 \theta_B, \\
\epsilon_{y y} & =S, \\
\epsilon_{z z} & =S \sin ^2 \theta_B+2 A \sin \theta_B \cos \theta_B+P \cos ^2 \theta_B, \\
\epsilon_{x y} & =-\epsilon_{y x}=i\left(D \cos \theta_B-C \sin \theta_B\right), \\
\epsilon_{x z} & =\epsilon_{z x}=A \cos 2 \theta_B+\left(S-P\right) \sin \theta_B \cos \theta_B, \\
\epsilon_{y z} & =-\epsilon_{z y}=-i\left(D \sin \theta_B+C \cos \theta_B\right).
\end{aligned}
\ee
The three electric components have
\be
E_z=-\epsilon_{z z}^{-1}\left(\epsilon_{z x} E_x+\epsilon_{z y} E_y\right).
\ee
Reinserting this back into the definition of electric displacement vector yields
\be
\left(\begin{array}{cc}
\eta_{x x}-n^2 & \eta_{x y} \\
\eta_{y x} & \eta_{y y}- n^2
\end{array}\right)\left(\begin{array}{c}
E_x \\
E_y
\end{array}\right)=0
\label{eq:deltaE}
\ee
where 
\be
\begin{aligned}
\eta_{x x} & =\frac{1}{\epsilon_{z z}}\left(\epsilon_{z z} \epsilon_{x x}-\epsilon_{x z} \epsilon_{z x}\right)=\frac{1}{\epsilon_{z z}}\left(SP -A^2\right) \\
\eta_{y y} & =\frac{1}{\epsilon_{z z}}\left(\epsilon_{z z} \epsilon_{y y}-\epsilon_{y z} \epsilon_{z y}\right) \\
& =\frac{1}{  \epsilon_{z z}}\left[\left(S^{2}-D^2-S  P +C^2\right) \sin ^2 \theta_B\right]\\
&+\frac{1}{  \epsilon_{z z}}\left[2\left(A S -C D\right) \sin \theta_B \cos \theta_B+S  P -C^2\right] \\
\eta_{y x} & =-\eta_{x y}=\frac{1}{\epsilon_{z z}}\left(\epsilon_{z z} \epsilon_{y x}-\epsilon_{y z} \epsilon_{z x}\right) \\
& =\frac{-i}{\epsilon_{z z}}\left[PD \cos \theta_B-S  C \sin \theta_B+A\left(D \sin \theta_B-C \cos \theta_B\right)\right]
\end{aligned}
\ee

For a normal pulsar or magnetar, one can obtain $\omega_B\gg\omega$ and $\omega_B\gg\omega_{\mathrm{p}}$ inside the magnetosphere.
Under these conditions, some parameters can be reduced as
\be
\begin{aligned}
& f_{11} \sim \omega_{{\rm p},s}^2\gamma_s/\omega_B^2 \sim 0, \quad f_{12} \sim 0,\\
&f_\eta \simeq-\omega_{{\rm p},s}^2\omega^{-2} \gamma_s^{-3}\left(1-n \beta_s \cos \theta_B\right)^{-2}, \\
& S \simeq 1, \quad P \simeq 1+f_\eta, \quad D \simeq 0, \quad A \simeq 0,\\
& C \simeq 0,\quad \epsilon_{z z} \simeq 1+f_\eta \cos ^2 \theta_B,\\
& \eta_{x x} \simeq \frac{1+f_\eta}{1+f_\eta \cos ^2 \theta_B},\quad\eta_{y y} \simeq 1 ,\quad\eta_{y x} \simeq-\eta_{x y} \simeq 0.
\end{aligned}
\ee
Reinserting this back into Equation (\ref{eq:deltaE}) and the solution of refractive index is given by \citep{1986ApJ...302..120A}
\be
\begin{aligned}
&n^2=1,\\
&n^2=\eta_{xx},\\
&\left(\omega^2-c^2 k_{\|}^2\right)\left[1-\frac{\omega_{{\rm p},s}^2}{\gamma_s^3\omega^2(1-\beta_s c k_{\|}/\omega)}\right]-c^2 k_{\perp}^2=0,
\end{aligned}
\label{eq:3solutions}
\ee
where $k_{\|}=k\cos\theta_B$ and $k_{\perp}=k\sin\theta_B$.
For the mode with $n^2=1$, the polarization is given by
\be
\left|\frac{E_x}{E_y}\right|=0, \quad \left|\frac{E_z}{E_y}\right|=0.
\ee
As a result, the mode is a transverse wave with the electric field vector in the $\boldsymbol{k}\times \boldsymbol{B}$ direction, that is, the X-mode.
For the other two refractive index solutions, one can obtain
\be
\left|\frac{E_y}{E_x}\right|=0, \quad \left|\frac{E_z}{E_x}\right|=\left|\frac{n^2-1}{\tan\theta_B}\right|.
\ee
The mode is called O-mode which the electric field oscillates at $\boldsymbol{k}-\boldsymbol{B}$ plane.

We plot the dispersion relationship of the X-mode and the O-mode for two cases ($\gamma=2$ and $\gamma=10$) in Figure \ref{fig:modes}. For both cases, $\theta_B=0.2$ is adopted.
The O-modes have two branches: the superluminous branch ($n<1$) and the subluminous branch ($n>1$).
The superluminous O-mode has a frequency cut-off at $\omega\gamma^3=\omega_{\mathrm{p}}$.
The latter one has a refractive index cut-off at $n=1/\cos\theta_B$.
This mode looks to have two branches as shown in Figure \ref{fig:modes}, but the branches can get crossed at high frequencies.

\section{Notation List}\label{app:notation}
Subscript $i$, $j$, $k$: The identifier of each charged particle\\
Subscript $s$: Species of a single particle
$c$: Speed of light\\
${\rm d}^2W/{\rm d}\omega {\rm d}\Omega$: Energy radiated per unit solid angle per unit frequency interval\\
$e$: Charge (absolute value) of electron\\
$f(u)$: Distribution function of plasma\\
$h_{\rm gap}$: Height of the gap\\
$k$: Wave vector\\
$k_{\mathrm{B}}$: Boltzmann constant\\
$l$: Longitude bunch size\\
$\ell$: Angular separation constant\\
$m_{\mathrm{e}}$: Mass of the electron\\
$\boldsymbol{n}$: Unit vector of the line of sight\\
$n$: Refractive index\\
$n_{\mathrm{e}}$: Number density of charges \\
$n_{\rm GJ}$: Goldreich-Julian density \\
$n_{\mathrm{p}}$: Number density of magnetospheric plasma\\
$p$: Power law index\\
$r$: Distance to the neutron star center\\
$s$: Distance that charges traveled along the field lines\\
$u$: Space-like part of the four-velocity\\
$u_m$: Average $u$ over the Gaussian distribution\\
$v$: Velocity of electron\\
$w$: Burst width\\
$A_{\parallel}$: Parallel component of amplitude\\
$A_{\perp}$: Perpendicular component of amplitude\\
$\boldsymbol{B}$: Magnetic field strength\\
$\boldsymbol{E}$: Wave electric field\\
$E_{1,\|}$: Perturbation electric field parallel to the $\boldsymbol{B}$-field\\
$E_{1,\perp}$: Perturbation electric field perpendicular to the $\boldsymbol{B}$-field\\
$E_{\parallel}$: Electric field parallel to the $\boldsymbol{B}$-field\\
$I,\,Q,\,U,\,V$: Stokes parameters\\
$\boldsymbol{J}$: Current density \\
$j$: Emissivity \\
$K_{\nu}$: Modified Bessel function \\
$\mathcal{L}_{\mathrm{b}}$: Luminosity of a bunch\\
$M$: Stellar mass\\
$\mathcal{M}$: Multiplicity factor \\
$N_{\mathrm{b}}$: Number of bunches that contribute to the observed power at an epoch \\
$N_{\mathrm{e}}$: Number of electrons in one bunch \\
$N_i$: Maximum number of the subscript of $i$\\
$N_j$: Maximum number of the subscript of $j$\\
$N_k$: Maximum number of the subscript of $k$\\
$P$: Spin period\\
$R$: Stellar radius\\
$R_{\rm b}$: Transverse size of bunch \\
$\mathcal{R}$: Distance from the emitting source to the observer \\
$T$: Temperature \\
$\mathcal{T}$: The average timescale of the flux density $I$ repeated\\
$Y_{\ell m} (\theta, \phi)$: Spherical harmonics\\
$\alpha$: Angle between spin axis and magnetic axis\\
$\beta$: Dimensionless velocity of one charge \\
$\dot{\beta}_\|$: Dimensionless perturbation acceleration parallel to the $\boldsymbol{B}$-field \\
$\dot{\beta}_\perp$: Dimensionless perturbation acceleration perpendicular to the $\boldsymbol{B}$-field\\
$\delta_V$: Vacuum polarization coefficient \\
$\gamma$: Lorentz factor\\
$\mu$: Shear modulus
$\nu$: Frequency, index of the modified Bessel function \\
$\dot\nu$: Drift rate\\
$\chi$: Angle between $\boldsymbol{\hat{\beta}}$ for different trajectories \\
$\chi_{d}$: Lower boundary of angle between $\boldsymbol{\hat{\beta}}$ for different trajectories \\
$\chi_{u}$: Upper boundary of angle between $\boldsymbol{\hat{\beta}}$ for different trajectories \\
$\boldsymbol{\epsilon}$: Dielectric tensor \\
$\lambda$: Wavelength\\
$\psi$: Polarization angle given by the RVM model\\
$\rho$: Curvature radius \\
$\boldsymbol{\sigma}$: Conductivity tensor \\
$\sigma_{\mathrm{c}}$: Curvature cross section \\
$\tau$: Optical depth \\
$\theta$: Poloidal angle with respect to the magnetic axis\\
$\theta_c$: Spread angle of the curvature radiation \\
$\theta_{\rm in}$: Angle between the incident photon momentum and the electron momentum\\
$\theta_s$: Angle of the footpoint for field line at the surface \\
$\theta_B$: Angle between angle between $\boldsymbol{\hat{k}}$ and $\boldsymbol{\hat{B}}$\\
$\omega$: Angular frequency \\
$\omega_a$: Frequency of the perturbation-induced acceleration \\
$\omega_B$: Cyclotron frequency \\
$\omega_{\mathrm{c}}$: Critical frequency of curvature radiation \\
$\omega_{\rm cut}$: Angular frequency cut-off of the superluminous O-mode \\
$\omega_{\rm in}$: Angular frequency of incoming wave of ICS \\
$\omega_M$: Frequency of bulk of bunches distribution \\
$\omega_{\mathrm{p}}$: Plasma frequency \\
$\varphi$: Azimuth angle to the magnetic
axis, the angle between LOS and the trajectory plane \\
$\varphi_{d}$: Lower boundary of angle between LOS and the trajectory plane \\
$\varphi_t$: Half opening angle of a bulk of bunches \\
$\varphi_{u}$: Upper boundary of angle between LOS and the trajectory plane \\
$\Gamma(\nu)$: Gamma function \\
$\Psi$: Polarization angle \\
$\Theta_e$: Magnetic colatitude of the emission point\\
$\Omega$: Solid angle of radiation, angular frequency of the oscillations.\\

\bibliographystyle{aa} 
\bibliography{ref} 

\end{document}